\documentclass[10pt,a4paper]{article}
\usepackage{amsmath}
\usepackage{amssymb}
\usepackage{graphics}
\usepackage{graphicx}
\title{Dynamical generation of fermion mass in a scalar-fermion theory
with $\lambda\phi^4$ interaction}

\author{Somnath Majumder and Krishnendu Mukherjee\\
Department of Physics, Indian Institute of Engineering Science and Technology,
Shibpur, \\Howrah-711103, West Bengal, India.}

\begin{document}

\maketitle

\begin{abstract}
The effective potential for a scalar theory with $\lambda\phi^4$ 
interaction, coupled to a massless fermion through Yukawa interaction 
is calculated by summing over infinite number of two particle
irreducible (2PI) diagrams of two different types and
a 2PI diagram of a third type using Cornwall, Jackiw
and Tomboulis (CJT) method.
There is an inversion symmetry present in the effective potential 
under $\phi\rightarrow -\phi$.
When the value of coupling constant falls beyond an open set
of positive real numbers, the effective potential exhibits
both maxima and minima above and below the zero potential line 
respectively on either side of its minimum at $\phi=0$.
The fermion acquires a mass in this region of 
coupling constant when the system 
settles into the minimum at positive, non-zero $\phi$  
breaking the inversion symmetry of the vacuum. 
However, the effective potential exhibits a minimum only at $\phi=0$
and also the fermion remains massless
when the coupling constant assumes any value from this open set.

\end{abstract}

\newpage

\tableofcontents

\section{Introduction}
\label{introduction}

It is always interesting to study the behaviour of quantum field theory
at large coupling which is commonly known as non-perturbative regime
of the theory. It is believed that the theory in this regime may behave
non-trivially which is completely absent in the perturbative 
expansion in powers of coupling constant to a finite order.
Attempts were made in this direction to understand the behaviour of
quantum field theories in the non-perturbative regime\cite{Jaffe1965,
Dashen1974}. The estimates using large order perturbation series
were also made for this purpose\cite{Lipatov1977}.
There was an investigation to produce symmetry breaking
generated by radiative corrections in gauge theories using
perturbative framework of field theory\cite{Coleman1973}.
There is a method of composite operators developed by Cornwall, Jackiw
and Tomboulis (CJT) \cite{Cornwall1974} to address
the non-perturbative issues in quantum field theories. This method has 
been employed to study non-perturbative effects on the phase 
transitions in scalar field theories \cite{Camelia1993, Camelia1996}.
This has also been used to study kaon condensation in colour-flavour
locked phase \cite{Alford2008, Andersen2009} and also in the linear 
sigma model at finite density and temperature \cite{Tran2008}.

In this paper we take a scalar field theory with $\lambda\phi^4$ interactions 
\cite{Agodi1998, Chung1997, Ananos2006} and the scalar is coupled
to a massless fermion field through Yukawa interaction.
We intend to generate the mass of fermion in this theory which itself is 
spontaneously unbroken in the classical level because 
the square of the mass parameter ($m^2$) of the scalar field and the 
coupling constant ($\lambda$) 
are assumed to be greater than zero. The fermion 
may acquire mass through Yukawa interactions if the effective potential 
possesses a minimum at non-zero value of $\phi$ ($=\langle 0\mid\Phi\mid
 0\rangle$). The effective potential may presumably exhibit this non-trivial 
behaviour if the loop corrections to infinite order are 
incorporated into it.
In Coleman-Weinberg model\cite{Coleman1973} the one-loop effective 
potential for $m^2=0$ exhibits a minimum at a very small non-zero $\phi$,
which occurs due to cancellation between the classical and the one-loop 
corrected terms. This violates the spirit of perturbation theory and is 
rectified by adopting the method of renormalization 
group\cite{Peskin and Schroeder}. The leading
logarithms to all orders of perturbation theory can be summed exactly
and the renormalization group improved effective potential exhibits 
no minimum at non-zero $\phi$. 
In this paper we shall
obtain the effective potential non-perturbatively by summing 
over a few classes of infinite set of diagrams that are 
summable to known algebraic form. 
It may so happen that the magnitude of the non-perturbative quantum 
corrections are comparable with the classical terms of the 
effective potential for certain range of values of the 
coupling constants of the theory. Then, it will not be quite 
unjustified if the cancellation occurs in this model between classical 
and the non-perturbative, quantum corrected terms for establishing 
minima of the potential at non-zero $\phi$.  

It is expected that the non-perturbative effects may lead to the 
appearance of non-trivial minima in the effective potential of this model and 
the consequent generation of fermion mass. 
The understanding of all these require the knowledge about the   
two point correlation functions of the fields present in the theory.
Therefore, it is a necessity to obtain a general formalism to 
derive equations for the non-perturbative, two point correlation functions 
of the theory and such method for field theory is  
developed by Cornwall-Jackiw-Tomboulis (CJT)\cite{Cornwall1974}.
We use this method to compute the effective potential $V_{eff}(\phi, G, S)$ 
as a functional of two-point correlation function between scalar 
fields ($G$) and the fermion fields ($S$) for constant $\phi$. 
Moreover, this effective potential is the generating functional for 
two-particle irreducible (2PI) Green's functions expressed in terms of 
the propagators $G$ and $S$ for a given $\phi$. In this method
$V_{eff}(\phi, G, S)$ is obtained as the sum of classical potential, 
one loop corrections and contributions from two and higher loop, 
2PI diagrams for constant $\phi$ denoted by $V_2(\phi, G, S)$.
We have identified two different types of 2PI diagrams each of which
containing infinite number of members contributing to $V_2(\phi, G, S)$
and the sum of them of either types can be obtained in closed form.
A third type of a two loop, 2PI diagram of order $\lambda$, 
which is absent in the first two types is also considered for its 
contribution to $V_2(\phi, G, S)$.
The extremization of the effective 
potential with respect to the independent variations of $G$ and $S$ 
gives rise to two gap equations that are self-consistently solved to 
obtain $G$ and $S$. The forms of $G$ and $S$ are then used to obtain the 
effective potential of the theory as a function of $\phi$.
Since the closed form expression of the effective potential 
is obtained after summing up of infinite set of 2PI 
diagrams, the result is inherently non-perturbative in nature.
In this paper we have considered only the diagrams of definite 
types, the sum of whose infinite members can be written in closed 
forms just to gain some preliminary insights about the behaviour of the 
system under the inclusion of such quantum corrections to order infinity.
We see that the constraint of two particle irreducibility 
makes it easier to identify the infinitely summable diagrams
belonging to different types. This is one of the reasons to use
the CJT method for incorporating non-perturbative corrections to the 
effective potential. This investigation may provide an idea about 
how a fraction of the total non-perturbative contribution can 
lead to spontaneous symmetry breaking and generation of fermion mass.

We observe that the effective potential exhibits maxima and minima
above and below the zero potential line respectively
at non-zero $\phi$ on either side of the minimum at $\phi=0$
in the following regions of the coupling constant:
$0 <\hat{\lambda}\le 0.32$ and $1.6\le\hat{\lambda}\le 3.0$
($\hat{\lambda}=\lambda/16\pi^2$).
Therefore, in this region of the coupling constant the system 
may settle in any one of the two minima appearing at non-zero 
$\phi$. The fermion coupled to scalar field acquires a mass if 
the system settles into the positive minimum and this 
breaks the inversion symmetry of the vacuum of the theory.
However, the effective potential exhibits a single minimum at $\phi=0$ 
when the coupling constant lies in the region
$0.32 <\hat{\lambda}< 1.6$, where the fermion remains massless.

The paper is organized as follows:   
The method of obtaining effective action
for composite operators proposed by
Cornwall-Jackiw-Tomboulis (CJT) is briefly discussed in Sec.\ref{cjtmethod}.
The effective potential for scalar- 
fermion theory is given in Sec.\ref{effectivepotential1}.
The methods of summing infinite number of 2-particle irreducible (2PI)
diagrams of two different types
and a single diagram of third type are discussed
in Sec.\ref{summablediagrams}.
The solution of the stationary state conditions of the effective
potential under the variation of scalar and fermion two point 
correlation functions are
discussed in Sec.\ref{delvdelg}. The effective potential and its
renormalization are discussed in Sec.\ref{effectivepotential}.
The discussion of the results are presented in
Sec.\ref{results and discussions}.
The evaluation of the integrals appeared in
Sec.\ref{delvdelg} are given
in Appendices \ref{IpJp}, \ref{dGamma2a}, \ref{dGamma2bc}.
The evaluation of the integrals appeared in Sec.\ref{effectivepotential}
are given in the Appendix \ref{veffintegrals}.

\section{Effective Action for Composite Operators using 
Cornwall-Jackiw-Tomboulis Method}
\label{cjtmethod}

We discuss briefly the method for obtaining effective potential 
for composite operators proposed by 
Cornwall-Jackiw-Tomboulis (CJT)\cite{Cornwall1974}.
Consider an action for scalar and fermion fields:
\begin{eqnarray}
	{\cal{I}}(\Phi,\bar{\psi},\psi) &=& \int d^4x {\cal{L}}(\Phi(x)
	,\bar{\psi(x)},\psi(x)).
\end{eqnarray}
$\Phi$ denotes a single component scalar field. $\psi$ and $\bar{\psi}$
denote a four component fermion field and its conjugate respectively.
Generating functional for Green's functions of nonlocal, composite fields is 
defined as
\begin{eqnarray}
	Z(J, K, \eta, \bar{\eta}, N) &=& e^{\frac{i}{\hbar}W(J,K,\eta,
	\bar{\eta},N)}\nonumber\\
	&=& \int {\cal{D}}\Phi {\cal{D}}\bar{\psi}{\cal{D}}\psi
	\exp\Big[\frac{i}{\hbar}\Big\{{\cal{I}}(\Phi,\bar{\psi},\psi)
	+ \int d^4x \Phi(x)J(x)
	+\frac{1}{2}\int d^4x d^4y \Phi(x)K(x,y)\Phi(y)\nonumber\\
	& &+\int d^4x \bar{\psi}(x)\eta(x)+\int d^4x \bar{\eta}(x)
	\psi(x)+\int d^4xd^4y \bar{\psi}(x)N(x,y)\psi(y)\Big\}\Big].
\end{eqnarray}	
$J$, $\eta$ and $\bar{\eta}$ are the sources for the fields 
$\Phi$, $\bar{\psi}$ and $\psi$ respectively. $K(x,y)$ and 
and $N(x,y)$ are the sources for the composite operators 
${\rm T}(\Phi(y)\Phi(x))$ and $-{\rm T}(\psi(y)\bar{\psi(x)})$ 
respectively.
We define
\begin{eqnarray}
	\frac{\partial W}{\partial J(x)} &=& <0|\Phi(x)|0> 
	= \phi(x),\nonumber\\
	\frac{\partial W}{\partial \eta(x)} &=&<0|\bar{\psi}(x)|0>=0,\nonumber\\
	\frac{\partial W}{\partial \bar{\eta}(x)} &=&<0|\psi(x)|0>=0,\nonumber\\
	\frac{\partial W}{\partial K(x,y)} &=& \frac{1}{2}
	\{\phi(y)\phi(x)+\hbar G(y,x)\},\nonumber\\
	\frac{\partial W}{\partial N_{\alpha,\beta}(x,y)} &=&-\hbar 
	S_{\beta,\alpha}(y,x).
\end{eqnarray}	
$\Gamma(\phi,G, S)$ is a double Legendre transform of 
$W(J, K, \eta, \bar{\eta}, N)$. 
We eliminate $J$, $K$ and $N$ in favour of $\phi$, $G$ and $S$ and set
\begin{equation}
	\Gamma (\phi,G,S) = W(J, K,\eta, \bar{\eta}, N)
	-\int d^4x \phi(x) J(x)
	-\frac{1}{2}\int d^4x d^4y \phi(x)K(x,y)\phi(y)
	- \frac{1}{2}\hbar{\rm Tr}GK + \hbar{\rm Tr}SN.
\end{equation}	
${\rm Tr}$ denotes the trace over space-time and internal space indices. 
Therefore,
\begin{eqnarray}
	\frac{\partial \Gamma (\phi,G,S)}{\partial \phi(x)} &=& -J(x)
	-\int d^4y K(x,y)\phi(y),
\label{dwrtphi1}\\
	\frac{\partial \Gamma (\phi,G,S)}{\partial G(x,y)} &=& -\frac{1}{2}
	\hbar K(y,x).
	\label{dwrtG1}\\
	\frac{\partial \Gamma (\phi,G,S)}{\partial S_{\alpha,\beta}(x,y)} &=& 
	\hbar N_{\beta, \alpha}(y,x).
        \label{dwrtS1}
\end{eqnarray}	
$\alpha, \beta$ ($= 1, \cdots 4$) denote the components of the 
fermion fields. Now,
\begin{eqnarray}
	e^{\frac{i}{\hbar}\Gamma(\phi,G,S)} 
	&=& \exp\Big[\frac{i}{\hbar}
	\Big\{ W-\int d^4x \phi(x) J(x)
        -\frac{1}{2}\int d^4x d^4y \phi(x)K(x,y)\phi(y)
	-\frac{1}{2}\hbar{\rm Tr}GK
	+ \hbar{\rm Tr}SN\Big\}\Big]\nonumber\\
	&=& \int {\cal{D}}\Phi {\cal{D}}\bar{\psi}{\cal{D}}\psi 
	\exp \Big[\frac{i}{\hbar}
	\Big\{{\cal{I}}(\Phi,\bar{\psi},\psi)+\int d^4x(\Phi(x)-\phi(x))J(x)
	+\int d^4x \bar{\psi}(x)\eta(x)\nonumber\\
	& &+\int d^4x \bar{\eta}(x)\psi(x)
	+\frac{1}{2}\int d^4x d^4y \Phi(x)K(x,y)\Phi(y) + \int d^4xd^4y
	\bar{\psi}(x)N(x,y)\psi(y)\nonumber\\
	& &-\frac{1}{2}\int d^4x d^4y 
	\phi(x)K(x,y)\phi(y)
	-\frac{1}{2}\hbar {\rm Tr} GK
	+\hbar{\rm Tr}SN\Big\}\Big].
\end{eqnarray}	
Replace $\Phi$ by $\Phi+\phi$,
\begin{eqnarray}
	e^{\frac{i}{\hbar}\Gamma(\phi,G,S)}
	&=& \int {\cal{D}}\Phi {\cal{D}}\bar{\psi}{\cal{D}}\psi
	\exp \Big[\frac{i}{\hbar}
	\Big\{{\cal{I}}(\Phi+\phi;\bar{\psi},\psi)+\int d^4x \Phi(x) 
	\Big(J(x)+\int d^4y K(x,y)\phi(y)\Big)\nonumber\\
	& &+ \frac{1}{2}\int d^4x d^4y \Phi(x)K(x,y)\Phi(y)
	+\int d^4x \bar{\psi}(x)\eta(x)+\int d^4x \bar{\eta}(x)\psi(x)
	+\int d^4xd^4y \bar{\psi}N(x,y)\psi(y)\nonumber\\
	& &-\frac{1}{2}\hbar {\rm Tr}GK
	+\hbar {\rm Tr} SN\Big\}\Big].
\end{eqnarray}	
Expand ${\cal{I}}(\Phi+\phi,\bar{\psi},\psi)$ around 
the point $c\equiv (\Phi=0, \bar{\psi}=0, \psi=0)$:
\begin{eqnarray}
	{\cal{I}}(\Phi+\phi)&=& {\cal{I}}(\phi)
	+\int d^4x \Phi(x)
	\frac{\partial{\cal{I}}}{\partial \phi(x)}\!\!\mid_c
	+\frac{1}{2} \int d^4xd^4y \Phi(x)
        \frac{\partial^2 {\cal{I}}}{\partial \phi(y)\partial 
	\phi(x)}\!\!\mid_c\Phi(y)
	+\int d^4x \bar{\psi}(x)
	\frac{\partial{\cal{I}}}{\partial \bar \psi(y)}\!\!\mid_c\nonumber\\
	& &+ \int d^4x \frac{\partial{\cal{I}}}{\partial \psi(x)}
	\!\!\mid_c\psi(x)
	+\int d^4xd^4y \bar{\psi}(x)
	\frac{\partial^2 {\cal{I}} }{\partial \psi(y)\partial 
	{\bar{\psi}}(x)}\!\!\mid_c\psi(y)
	+{\cal{I_{\rm{int}}}}(\phi; \Phi,\bar{\psi},\psi).
\end{eqnarray}	
Therefore,
\begin{eqnarray}
	e^{\frac{i}{\hbar}\Gamma(\phi,G,S)}
	&=& \int {\cal{D}}\Phi {\cal{D}}\bar{\psi}{\cal{D}}\psi
	\exp \Big[\frac{i}{\hbar}
	\Big\{{\cal{I}}(\phi)+\int d^4x \Phi(x) \Big(\frac{
	\partial {\cal{I}}}{\partial \phi(x)}\!\!\mid_c
	+J(x)+\int d^4y K(x,y)\phi(y)\Big)\nonumber\\
	& &+ \frac{1}{2}\int d^4x d^4y \Phi(x)\Big(K(x,y)
	+\frac{\partial^2 {\cal{I}}}{\partial \phi(y)\partial
	\phi(x)}\!\!\mid_c\Big)\Phi(y)
	+\int d^4x \bar{\psi}(x)\Big(\frac{\partial {\cal{I}}}
	{\partial \bar{\psi}(x)}\!\!\mid_c+\eta(x)\Big)\nonumber\\
	& &+ \int d^4x \Big(\frac{\partial {\cal{I}}}
	{\partial\psi(x)}\!\!\mid_c+\bar{\eta}(x)\Big)\psi(x)
	+ \int d^4xd^4y \bar{\psi}(x)\Big(N(x,y)
	+\frac{\partial^2 {\cal{I}}}{\partial \psi(y)\partial{\bar{\psi}}(x)}
	\!\!\mid_c\Big)\psi(x)\nonumber\\
	& &-\frac{1}{2}\hbar{\rm Tr}GK+\hbar{\rm Tr}SN
	+{\cal{I_{\rm{int}}}}(\phi; \Phi,\bar{\psi},\psi)\Big\}\Big]
\end{eqnarray}	
Using the definitions
\begin{eqnarray}
	i{\cal{D}}^{-1}(\phi;x,y) &=& \frac{\partial^2 {\cal{I}}}
	{\partial \phi(y)\partial\phi(x)}\!\!\mid_c,\\
	i{\cal{S}}^{-1}(\phi;x,y) &=& \frac{\partial^2 {\cal{I}}}
	{\partial\psi(y)\partial\bar{\psi}(x)}\!\!\mid_c
\end{eqnarray}	
and the following equations of motion:
\begin{eqnarray}
	\frac{\partial {\cal{I}}}{\partial \phi(x)}\mid_c+J(x)+
	\int d^4y K(x,y) \phi(y)&=& 0,\nonumber\\
		\frac{\partial \cal{I}}{\partial \bar{\psi}(x)}\mid_c
		+\eta(x) &=& 0,\nonumber\\
		\frac{\partial \cal{I}}{\partial \psi(x)}\mid_c
		+\bar{\eta}(x) &=& 0.
\end{eqnarray}	
we obtain the effective action as
\begin{equation}
	\Gamma (\phi,G,S) = {\cal{I}}(\phi)-\frac{\hbar}{2}{\rm Tr}GK
	+\hbar{\rm Tr}SN-i\hbar \ln \int {\cal{D}}\Phi {\cal{D}}\bar{\psi}
	{\cal{D}}\psi \exp\Big\{
		\frac{i}{\hbar}{\cal{I}}(\phi,G,S;\Phi,\bar{\psi},
		\psi)\Big\}.
\label{ea1}
\end{equation}	
Here,
\begin{eqnarray}
	{\cal{I}}(\phi,G,S;\Phi,\bar{\psi},\psi)&=&
	\frac{1}{2}\int d^4xd^4y \Phi(x)
	\Big(K(x,y)+i{\cal{D}}^{-1}(\phi;x,y)\Big)\Phi(y)\nonumber\\
	& &+\int d^4xd^4y
	\bar{\psi}(x)\Big(N(x,y)+i{\cal S}^{-1}(\phi;x,y)\Big)\psi(x)
	+{\cal{I_{\rm{int}}}}(\phi;\Phi,\bar{\psi},\psi).
\end{eqnarray}
First we differentiate the Eq.(\ref{ea1}) with respect to $G(z,w)$ and obtain
\begin{eqnarray}
        \frac{\partial \Gamma}{\partial G(z,w)} &=& -\frac{\hbar}{2}
        \int d^4x d^4y \delta^4 (x-z)\delta^4 (y-w) K(y,x)
	-\frac{\hbar}{2}\int d^4xd^4y 
	G(x,y)\frac{\partial K(y,x)}{\partial G(z,w)}\nonumber\\
	& &-i\hbar\frac{\int{\cal{D}}\Phi{\cal{D}}\bar{\psi}{\cal{D}}\psi 
	e^{\frac{i}{\hbar}{\cal{I}}(\phi,G,S;\Phi,\bar{\psi},\psi)}
	\frac{i}{2\hbar}\int d^4xd^4y \Phi(y)
	\frac{\partial K(y,x)}{\partial G(z,w)}\Phi(x)}
	{\int {\cal{D}}\Phi{\cal{D}}\bar{\psi}{\cal{D}}\psi 
	e^{\frac{i}{\hbar}{\cal{I}}(\phi,G,S;\Phi,\bar{\psi},\psi)}}.
\end{eqnarray}
It gives
\begin{equation}
\hbar G(x,y) = \frac{\int{\cal{D}}\Phi{\cal{D}}\bar{\psi}{\cal{D}}\psi 
	e^{\frac{i}{\hbar}{\cal{I}}(\phi,G,S;\Phi,\bar{\psi},\psi)}
	\Phi(x)\Phi(y)}
	{\int {\cal{D}}\Phi{\cal{D}}\bar{\psi}{\cal{D}}\psi
	e^{\frac{i}{\hbar}{\cal{I}}(\phi,G,S;\Phi,\bar{\psi},\psi)}}.
	\label{defG}
\end{equation}
Next, differentiating Eq.(\ref{ea1}) with respect to $S_{\alpha,\beta}(z,w)$
we obtain
\begin{eqnarray}
	\frac{\partial \Gamma}{\partial S_{\alpha,\beta}(z,w)} &=&
	\hbar N_{\beta,\alpha}(w,z) + \hbar \int d^4xd^4y S_{{\alpha^{\prime}}
	,\beta^{\prime}}(x,y)\frac{\partial N_{\beta^{\prime}\alpha^{\prime}}
	(y,x)}{\partial S_{\alpha,\beta}(z,w)}\nonumber\\
	& &-i\hbar\frac
	{\int {\cal{D}}\Phi {\cal{D}}\bar{\psi}
	{\cal{D}}\psi \frac{i}{\hbar}\int d^4x d^4y \bar{\psi}_{\beta^{\prime
	}}(y)\frac{\partial N_{\beta^{\prime}\alpha^{\prime}}
        (y,x)}{\partial S_{\alpha,\beta}(z,w)}\psi_{\alpha^{\prime
        }}(x)\exp\Big[\frac{i}{\hbar}{\cal{I}}(\phi,G,S;\Phi,\bar{\psi},\psi
	)\Big]}
	{\int{\cal{D}}\Phi {\cal{D}}\bar{\psi}{\cal{D}}\psi
        \exp\Big[\frac{i}{\hbar}{\cal{I}}(\phi,G,S;\Phi,\bar{\psi},\psi
        )\Big]} .
\end{eqnarray}	
It gives
\begin{equation}
	\hbar S_{\alpha^{\prime}\beta^{\prime}}(x,y) = 
	\frac{\int{\cal{D}}\Phi {\cal{D}}\bar{\psi} {\cal{D}}\psi
	\psi_{\alpha^{\prime}}(x) \bar{\psi}_{\beta^{\prime
        }}(y)e^{\frac{i}{\hbar}{\cal{I}}
	(\Phi,\bar{\psi},\psi;\phi,G,S)}}{\int{\cal{D}}\Phi {\cal{D}}
	\bar{\psi} {\cal{D}}\psi e^{\frac{i}{\hbar}{\cal{I}}
        (\Phi,\bar{\psi},\psi;\phi,G,S)}}.
	\label{defS}
\end{equation}
Eq.(\ref{defG}) and (\ref{defS}) suggest that $G$ and $S$ act 
as the exact connected propagators of the theory and therefore, 
the following relations must hold:
\begin{eqnarray}
	K(x,y) &=& iG^{-1}(x,y)-i{\cal D}^{-1}(\phi; x, y),\\
	N(x,y) &=& iS^{-1}(x,y)-i{\cal{S}}^{-1}(\phi; x, y).
\end{eqnarray}	
Then, the effective action takes the form
\begin{equation}
        \Gamma (\phi,G,S)={\cal{I}}(\phi)-\frac{\hbar}{2}Tr G\Big(iG^{-1}
	-i{\cal{D}}^{-1}(\phi)\Big)+\hbar Tr S\Big(iS^{-1}-i{\cal{S}}^{-1}
	(\phi)\Big) 
	-i\hbar \ln \int {\cal{D}}\Phi{\cal{D}}
        \bar{\psi} {\cal{D}}\psi \exp\Big\{\frac{i}{\hbar}{\cal{I}}
	(\phi,G,S;\Phi,\bar{\psi},\psi)\Big\},
\label{ea2}
\end{equation}
where,
\begin{equation}
	{\cal{I}}(\phi,G,S;\Phi,\bar{\psi},\psi)=\frac{1}{2}\int d^4xd^4y
	\Phi(x) iG^{-1}(x,y)\Phi(y) +\int d^4xd^4y
	\bar{\psi}(x) iS^{-1}(\phi;x,y)\psi(x) +{\cal{I_{\rm{int}}}}
	(\phi;\Phi,\bar{\psi},\psi).
\end{equation}
Now,
\begin{eqnarray}
	& &	\int {\cal{D}}\Phi{\cal{D}}
        \bar{\psi} {\cal{D}}\psi \exp\Big\{\frac{i}{\hbar}{\cal{I}}
	(\phi,G,S;\Phi,\bar{\psi},\psi)
	\Big\}\nonumber\\
	&=& \int {\cal{D}}\Phi \exp\Big\{\frac{i}{2\hbar}\int d^4xd^4y
	\Phi(x)iG^{-1}(x,y)\Phi(y)\Big\}\Big\{{\cal{D}}
        \bar{\psi} {\cal{D}}\psi\exp\Big\{\frac{i}{\hbar}
	\int d^4xd^4y\bar{\psi}(x)iS^{-1}(\phi;x,y)\psi(y) \Big\}\nonumber\\ 
	& &\times \frac{\int {\cal{D}}\Phi{\cal{D}}
        \bar{\psi} {\cal{D}}\psi 
	\exp\Big\{\frac{i}{\hbar}{\cal{I}}(\phi,G,S;\Phi,\bar{\psi},\psi)
        \Big\}}{\int {\cal{D}}\Phi{\cal{D}}
        \bar{\psi} {\cal{D}}\psi \exp\Big\{\frac{i}{2\hbar}\int d^4xd^4y
	\Phi(x)iG^{-1}(x,y)\Phi(y)+\frac{i}{\hbar}\int d^4xd^4y
	\bar{\psi}(x)iS^{-1}(\phi;x,y)\psi(y)\Big\}}\nonumber\\
	&=& \Big[{\rm Det}(iG^{-1})\Big]^{-\frac{1}{2}}
        \Big[{\rm Det}(iS^{-1})\Big]
	\exp\Big\{\frac{i}{\hbar}\Gamma_{2}(\phi,G,S)\Big\}.
\end{eqnarray}	
Therefore, the Eq.(\ref{ea2}) becomes
\begin{equation}
	\Gamma(\phi,G,S) = {\cal{I}}(\phi)
	+\frac{i\hbar}{2}{\rm TrLn} (iG^{-1})
	-i\hbar{\rm TrLn}(iS^{-1})
	+\frac{i\hbar}{2}{\rm Tr}G{\cal{D}}^{-1}
	-i\hbar{\rm Tr}S{\cal{S}}^{-1} + C
	+\Gamma_{2}(\phi,G,S)
	\label{ea3}
\end{equation}	
where $C$ is a $\phi$ independent constant and 
\begin{eqnarray}
	\exp\Big\{\frac{i}{\hbar}\Gamma_{2}(\phi,G,S)\Big\} &=&
\frac{\int {\cal{D}}\Phi{\cal{D}}
        \bar{\psi} {\cal{D}}\psi 
        \exp\Big\{\frac{i}{\hbar}{\cal{I}}(\phi,G,S;\Phi,\bar{\psi},\psi)
        \Big\}}{\int {\cal{D}}\Phi{\cal{D}}
        \bar{\psi} {\cal{D}}\psi \exp\Big\{\frac{i}{2\hbar}\int d^4xd^4y
        \Phi(x)iG^{-1}(x,y)\Phi(y)+\frac{i}{\hbar}\int d^4xd^4y
        \bar{\psi}(x)iS^{-1}(\phi;x,y)\psi(y)\Big\}}
\end{eqnarray}
$\Gamma_2(\phi,G,S)$ denotes the sum of all two particle irreducible 
vacuum graphs in the theory governed by the action 
${\cal{I}}(\phi,G,S;\Phi,\bar{\psi},\psi)$ and the propagators $G(x,y)$
and $S(x,y)$.
In the stationary state of the system 
we need to set the external sources $J$, $K$ and $N$ in 
Eq.(\ref{dwrtphi1}), (\ref{dwrtG1}) and (\ref{dwrtS1}) 
simultaneously to zero and obtain the following conditions:
\begin{eqnarray}
\frac{\partial\Gamma (\phi,G,S)}{\partial \phi(x)} &=& 0,
\label{dwrtphi2}\\
\frac{\partial\Gamma (\phi,G,S)}{\partial G(x,y)} &=& 0.
\label{dwrtG2}\\
\frac{\partial\Gamma (\phi,G,S)}{\partial S(x,y)} &=& 0.
\label{dwrtS2}
\end{eqnarray}
When $\phi$ is a constant, independent of $x$, the solutions of 
Eq.(\ref{dwrtphi2}) provides the extrema of the 
effective potential. Next, to find the physical significance of 
Eq.(\ref{dwrtG2}) and (\ref{dwrtS2}), we differentiate 
separately Eq.(\ref{ea3}) 
with respect to $G$ and $S$ and set the results to zero:
\begin{eqnarray}
	G^{-1}(x,y) &=& {\cal{D}}^{-1}(\phi;x, y) 
	- \Sigma_s(\phi, G, S; x, y),
\label{ssdeqn}\\
	S^{-1}(x,y) &=& {\cal{S}}^{-1}(\phi; x, y) 
	- \Sigma_f(\phi, G, S; x, y),
\label{fsdeqn}
\end{eqnarray}
where,
\begin{eqnarray}
	\Sigma_s(\phi, G, S; x, y) &=& \frac{2i}{\hbar}
\frac{\partial\Gamma_2(\phi, G, S)}{\partial G(x,y)},
	\label{sselfenergy}\\
	\Sigma_f(\phi, G, S; x, y) &=& -\frac{i}{\hbar}
\frac{\partial\Gamma_2(\phi, G, S)}{\partial S(x,y)},
	\label{fselfenergy}
\end{eqnarray}
Here, $\Sigma_s$ and $\Sigma_f$ denote the self-energy parts of the propagators 
$G$ and $S$ respectively. The Eq.(\ref{ssdeqn}) and (\ref{fsdeqn}) are 
called the Schwinger-Dyson equations for bosonic and fermionic  
propagators of the theory. 
Since self energy functions are obtained as a derivative of 
$\Gamma_2$ with respect to the propagators, 
$\Gamma_2$ is ensured to be a 2PI Green function of the theory.

\section{Scalar-Fermion theory with $\lambda\Phi^4$ interaction}
\label{effectivepotential1}

The theory is described by the action
\begin{equation}
	{\cal{I}}(\Phi, \psi, \bar{\psi})
=\int d^4x [\frac{1}{2}\partial_\mu\Phi\partial^\mu\Phi
-\frac{1}{2}m^2\Phi^2-\frac{\lambda}{4!}\Phi^4
+ \bar{\psi}(i\partial\!\!\!/ - g\Phi)\psi].
\end{equation}
We shift the field $\Phi$ by $\phi(x)$ and obtain the action as
\begin{equation}
{\cal{I}}(\Phi, \phi, \psi, \bar{\psi}) 
= {\cal{I}}(\phi)+\frac{1}{2}\int d^4x\Phi(x) 
i{\cal{D}}^{-1}(\phi(x); x, y)\Phi(y)
+\int d^4x\bar{\psi}i{\cal{S}}^{-1}(\phi(x); x, y)\psi(y)
+{\cal{I}}_{\rm{int}}(\Phi, \phi),
\end{equation}
where the equation of motion is used to get rid of the linear terms in
$\Phi(x)$. Here, $i{\cal{D}}^{-1}(\phi(x); x, y) = 
\frac{\partial^2{\cal{I}}}
{\partial\Phi(x)\partial\Phi(y)}\!\!\mid_c
= -[\Box_x+m^2+\frac{\lambda}{2}\phi(x)^2]\delta^{(4)}(x-y)$,
$i{\cal{S}}^{-1}(\phi(x); x, y)=
\frac{\partial^2{\cal{I}}}
{\partial\psi(y)\partial\bar{\psi}(x)}\!\!\mid_c
= (i\partial\!\!\!/ - g\phi)\delta^{(4)}(x-y)$
and the interaction term takes the form:
\begin{equation}
{\cal{I}}_{\rm{int}}(\Phi,\phi)=-\int d^4x [\frac{\lambda}{3!}\phi(x)\Phi^3(x)
+\frac{\lambda}{4!}\Phi^4(x)].
\end{equation}
The effective action for the composite operators which are taken as
scalar and fermion two point correlation functions 
$G(x,y)=\langle 0\mid T(\Phi(x)\Phi(y))\mid 0\rangle$ and  
$S(x,y)=\langle 0\mid T({\psi(x)\bar{\psi}}(y))\mid 0\rangle$ respectively
is written following Eq.(\ref{ea3}) as
a functional of $\phi$, $G$ and $S$:
\begin{equation}
	\Gamma(\phi,G,S)={\cal{I}(\phi)}
	+\frac{i}{2}{\rm Tr}{\rm Ln}(iG^{-1})
	+\frac{i}{2}{\rm Tr}({\cal{D}}^{-1}(\phi)G)
	-i{\rm Tr}{\rm Ln}(iS^{-1})-i{\rm Tr}({\cal{S}}^{-1}(\phi)S)
+\Gamma_2(\phi, G),
\end{equation}
where we set $\hbar =1$ and $C=0$.
$\Gamma_2(\phi, G)$ includes contributions of 2PI, two and more 
than two loop diagrams. When $\phi={\rm constant}$, the 
renormalized effective potential for a translation invariant theory reads as
\begin{eqnarray}
V_{eff}(\phi, G, S) &=& V(\phi)+V_{ct}(\phi)
-\frac{i}{2}\int\frac{d^4p}{(2\pi)^4}\ln\det G^{-1}(p)
        -\frac{i}{2}\int\frac{d^4p}{(2\pi)^4}
\{{\cal{D}}^{-1}(\phi, p)+G_{ct}^{-1}\}G(p)\nonumber\\
	& & +i\int\frac{d^4p}{(2\pi)^4}\ln\det S^{-1}(p)
+i \int\frac{d^4p}{(2\pi)^4}
{\cal{S}}^{-1}(\phi, p)S(p)
        +V_2(\phi, G),
\label{Veff1}
\end{eqnarray}
where the classical potential
\begin{equation}
        V(\phi)=\frac{1}{2}m^2\phi^2
+ \frac{\lambda}{4!}\phi^4.
\end{equation}
$V_2(\phi, G)$ contains contributions form two and higher than 
two loop, 2PI,  diagrams contributing to the effective potential and
\begin{equation}
        i{\cal{D}}^{-1}(p) = p^2-M^2(\phi),
\end{equation}
\begin{equation}
	i{\cal{S}}^{-1}(p) = p\!\!\!/ - M_f(\phi),
\end{equation}
where
\begin{eqnarray}
	M^2(\phi) &=& m^2+\frac{\lambda}{2}\phi^2,\\
	M_f(\phi) &=& g\phi.
\end{eqnarray}
We adopt the requirements following Eq.(\ref{dwrtG2}) and (\ref{dwrtS2}) 
that the effective potential is stationary with respect to the independent
variation of $G(p)$ and $S(p)$:
\begin{eqnarray}
	\frac{\partial V_{eff}(\phi, G, S)}{\partial G(p)} &=& 0,
\label{Geqn1} \\
	\frac{\partial V_{eff}(\phi, G, S)}{\partial S(p)} &=& 0.
\label{Seqn1}
\end{eqnarray}
The counter terms $V_{ct}(\phi)$ and $G_{ct}^{-1}$ have been 
introduced to remove ultraviolet divergences present in Eq.(\ref{Veff1})
and (\ref{Geqn1}) respectively. We shall discuss about the solution of the
equation for $G(p)$, $S(p)$ and the evaluation of the renormalized
effective potential using their solutions in the following sections.

\section{Diagrams contributing to $V_2(\phi, G)$}
\label{summablediagrams}

The 2PI diagrams of more than one loop contributing to $V_2(\phi, G)$
are classified in terms of their geometry into different types and 
each type may contain infinite number of members. The $n$-th member 
of a type possesses $n$ loops, where $n\ge 2$. We have identified 
so far two different types denoted by $a$ and $b$, 
where the sum of the infinite number of members of each type 
contributing to $V_2(\phi, G)$ is obtained in terms of known 
algebraic function. We have also identified a 2PI, two-loop diagram
whose geometry is different from those of members present 
in type-$a$ and $b$ and created a type denoted by $c$ containing this 
lone diagram as its member.
It is noteworthy that the constraint of two particle irreducibility 
provides a scope to identify diagrams of type-$a$ and $b$ which is 
one of the key advantages of using the CJT method.
It is true that there is a long list of diagrams geometrically 
belonging to types different from $a$, $b$ and $c$ have not 
been taken into consideration owing to our inability to obtain
their infinite sums in terms of known algebraic functions.
It means that we have been able to take account only a fraction of the 
total non-perturbative corrections contained for the given theory. 
This work may provide us a flavour of how the non-perturbative 
effects, whatever small it may be, bring about a change in our 
common understanding of field theory.

\begin{figure}[!th]
\begin{center}
\includegraphics[scale=0.4]{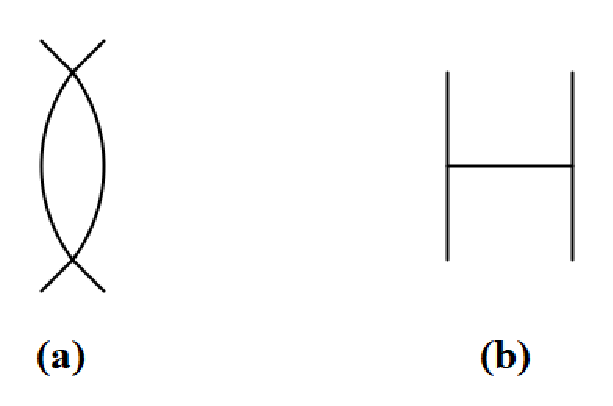}
\end{center}
\vskip -0.35in
        \caption[]{ (a) Lobe (b) Trilinear line }
\label{loli}
\end{figure}

\begin{figure}[!ht]
\begin{center}
\includegraphics[scale=0.12]{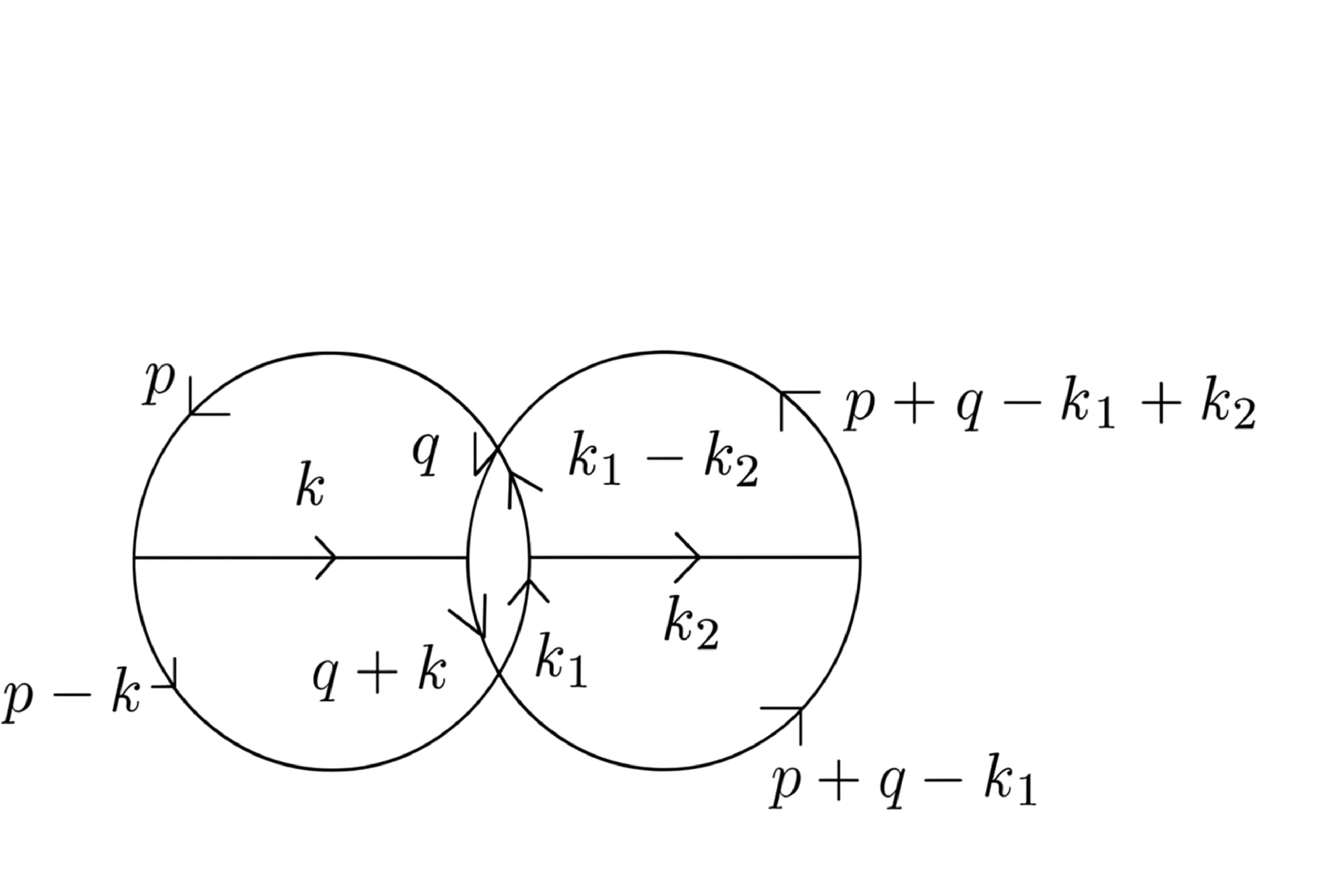}
\end{center}
\vskip -0.3in
        \caption[]{2PI diagram with two lobes each
        containing one trilinear line}
\label{lo2li2}
\end{figure}

\begin{figure}[!ht]
\begin{center}
\includegraphics[scale=0.35]{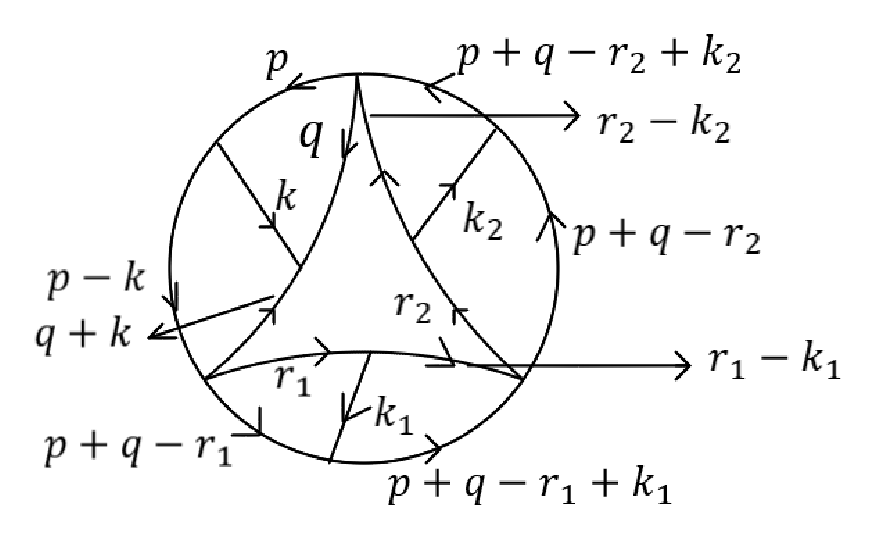}
\end{center}
\vskip -0.3in
        \caption[]{2PI diagram with three lobes each
        containing one trilinear line}
\label{lo3li3}
\end{figure}

\begin{figure}[!ht]
\begin{center}
\includegraphics[scale=0.3]{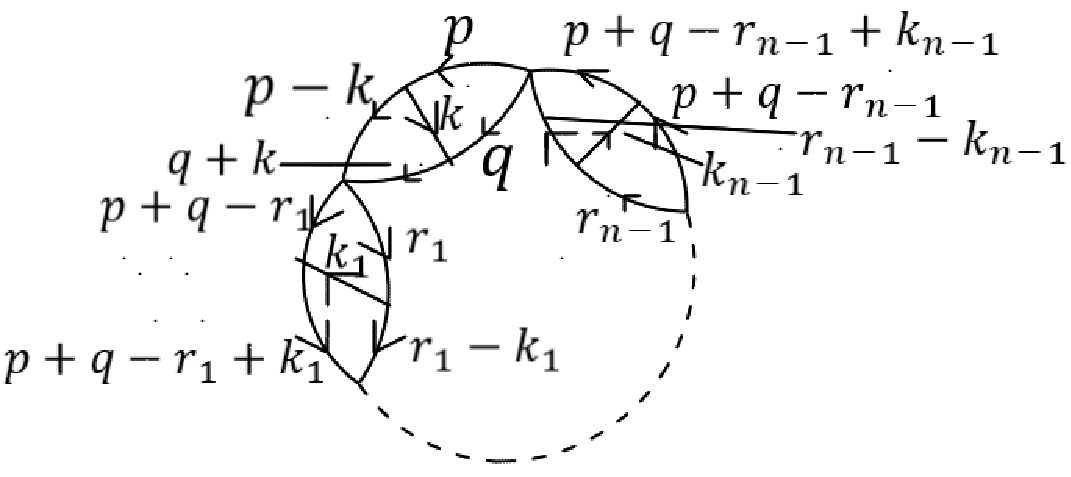}
\end{center}
\vskip -0.3in
        \caption[]{2PI diagram with n lobes each
        containing one trilinear line}
\label{lonlin}
\end{figure}

\begin{figure}[!ht]
\begin{center}
\includegraphics[scale=0.1]{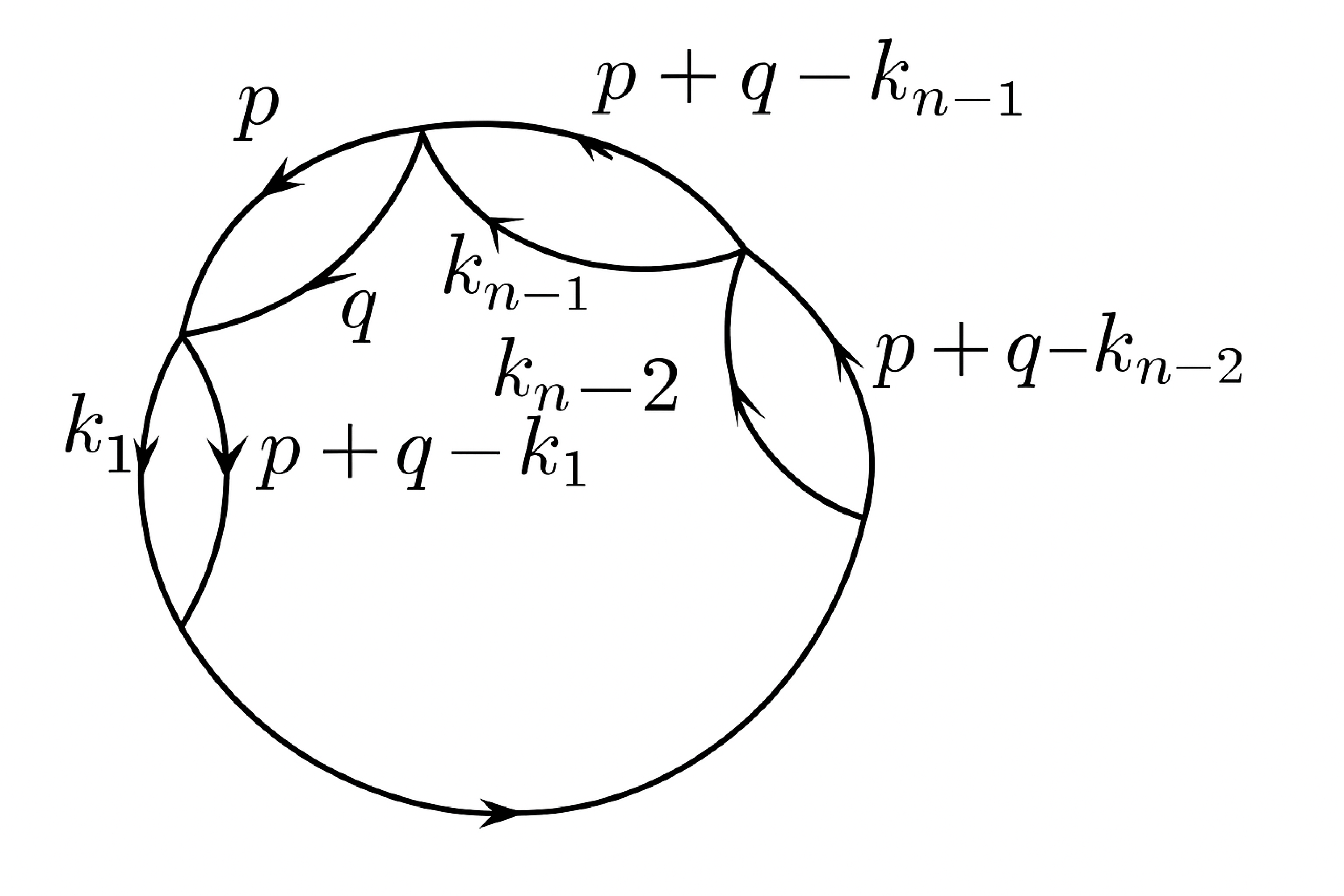}
\end{center}
\vskip -0.3in
        \caption[]{2PI diagram with n lobes
        }
\label{lonli0}
\end{figure}

\begin{figure}[!ht]
\begin{center}
\includegraphics[scale=0.4]{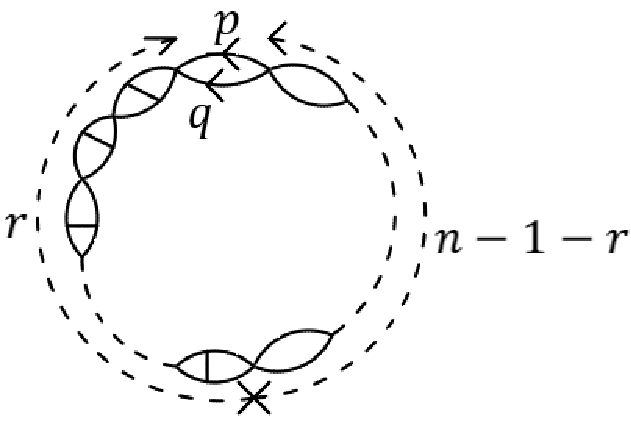}
\end{center}
\vskip -0.3in
        \caption[]{2PI diagram with n lobes
and $r$ trilinear lines}
\label{lonlir}
\end{figure}

\begin{figure}[!ht]
\begin{center}
\includegraphics[scale=0.1]{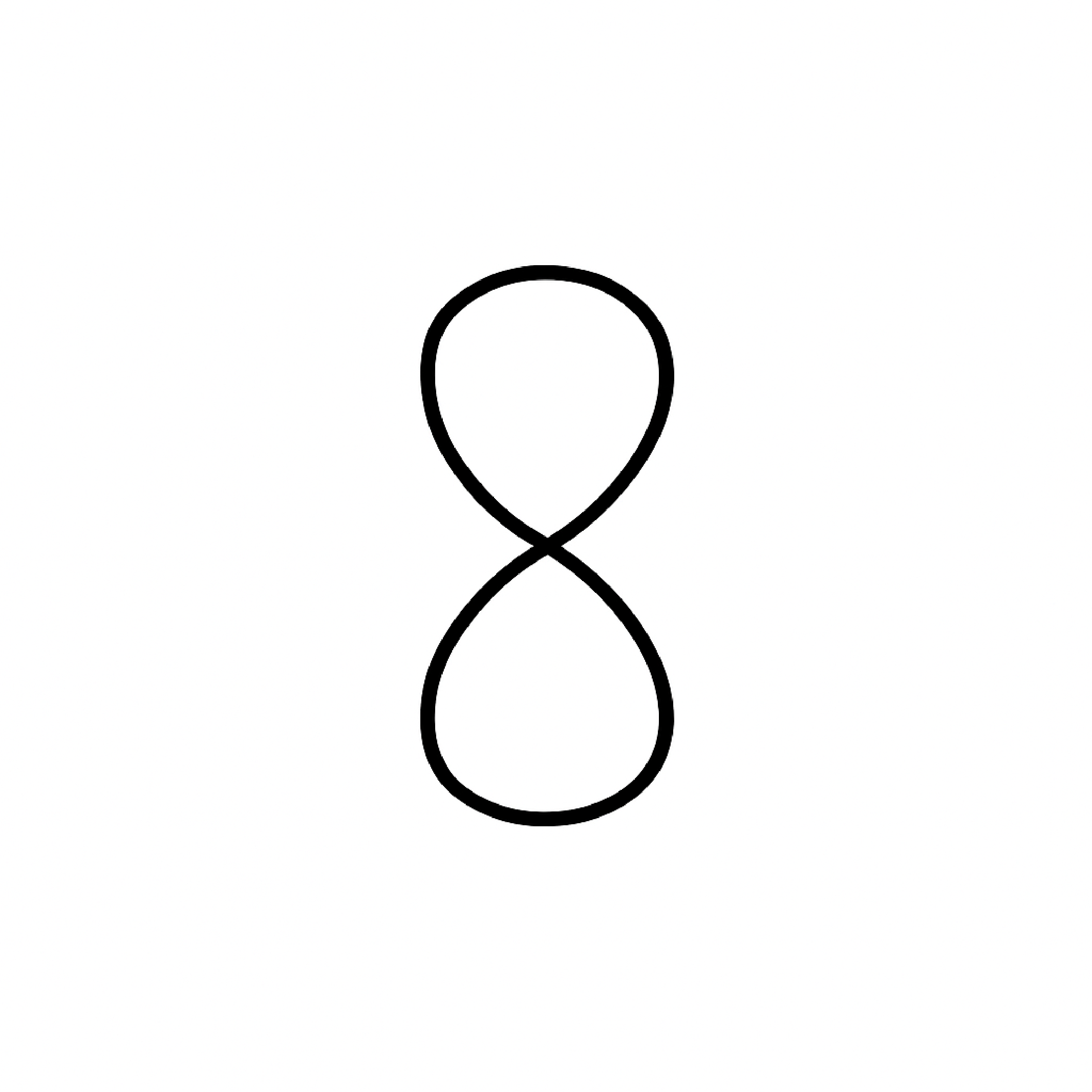}
\end{center}
\vskip -0.3in
        \caption[]{2PI diagram of type $c$.}
\label{g2c}
\end{figure}

We need to define some basic elements for this purpose. First,
we define a `lobe' which is constructed by joining any two lines
of a four point vertex to two lines of an another four point
vertex. Next, we also define a `trilinear line' which is constructed by
joining a line of a three point
vertex to a line of an another three point vertex. A lobe and a
a trilinear line are shown in FIG.\ref{loli}(a) and
\ref{loli}(b) respectively.
We shall first discuss about 2PI diagrams of a particular kind which
we name it as type-$a$.
A typical diagram of this kind is shown in FIG.\ref{lo2li2}
which involves two lobes each containing one trilinear line.
The contribution reads as
\begin{eqnarray}
	\gamma_a^{(2,2)} &=& \left(\frac{-i\lambda}{4!}\right)^2
\left(\frac{-i\lambda\phi}{3!}\right)^4
        \int \frac{d^4p}{(2\pi)^4}
        \frac{d^4q}{(2\pi)^4}
        \frac{d^4k}{(2\pi)^4}
        \frac{d^4k_1}{(2\pi)^4}
        \frac{d^4k_2}{(2\pi)^4}
        3G(p)3G(q)2G(k)4G(p-k)3G(q+k)
        3G(k_1)\nonumber\\
        & &\times 3G(p+q-k_1)2G(k_2)
        2G(k_1-k_2)
        G(p+q-k_1+k_2)\nonumber\\
        &=&\frac{1}{6}\left(\frac{i\lambda^3\phi^2}{4}\right)^2 
        \int \frac{d^4p}{(2\pi)^4}
        \frac{d^4q}{(2\pi)^4}
        \frac{d^4k}{(2\pi)^4}
        G(p)G(q)G(p-k)
        G(q+k)G(k)J(p+q)
\end{eqnarray}
where
\begin{equation}
        J(p) = \int\frac{d^4k_1}{(2\pi)^4}\frac{d^4k_2}{(2\pi)^4}
        G(k_1)G(p-k_1)G(k_1-k_2)G(p-k_1+k_2)G(k_2).
        \label{Jp}
\end{equation}
The contribution of the diagram in FIG.\ref{lo3li3} reads as
\begin{eqnarray}
	\gamma_a^{(3,3)} &=& \left(\frac{-i\lambda}{4!}\right)^3
\left(\frac{-i\lambda\phi}{3!}\right)^6
        \int \frac{d^4p}{(2\pi)^4}
        \frac{d^4q}{(2\pi)^4}
        \frac{d^4k}{(2\pi)^4}
        \frac{d^4r_1}{(2\pi)^4}
        \frac{d^4k_1}{(2\pi)^4}
        \frac{d^4r_2}{(2\pi)^4}
        \frac{d^4k_2}{(2\pi)^4}
        3G(p)3G(q)2G(k)4G(p-k)3G(q+k)\nonumber\\
        & &\times 3G(r_1)3G(p+q-r_1)4G(r_1-k_1)
        3G(p+q-r_1+k_1)2G(k_1)
        3G(r_2)3G(p+q-r_2)
        4G(r_2-k_2)\nonumber\\
	& &\times G(p+q-r_2+k_2)2G(k_2)\nonumber\\
        &=&\frac{1}{6}\left(\frac{i\lambda^3\phi^2}{4}\right)^3 
        \int \frac{d^4p}{(2\pi)^4}
        \frac{d^4q}{(2\pi)^4}
        \frac{d^4k}{(2\pi)^4}
        G(p)G(q)G(p-k)
        G(q+k)G(k)J^2(p+q)
\end{eqnarray}
Therefore, the contribution of the diagram in FIG.\ref{lonlin} involving
$n$ lobes each containing a trilinear line is given as
\begin{equation}
        \gamma_a^{(n,n)}=
\frac{1}{6}\left(\frac{i\lambda^3\phi^2}{4}\right)^n 
        \int \frac{d^4p}{(2\pi)^4}
        \frac{d^4q}{(2\pi)^4}
        \frac{d^4k}{(2\pi)^4}
        G(p)G(q)G(p-k)G(q+k)G(k)J^{n-1}(p+q)
\end{equation}
After summing all possible diagrams of type-$a$, we obtain their
contributions to the effective potential as
\begin{equation}
-iV^{(a)}_2(\phi, G) = \sum_{n=2}^{\infty}
\gamma_a^{(n,n)}
        =-\frac{\lambda^6\phi^4}{96}
\int \frac{d^4p}{(2\pi)^4}
        \frac{d^4q}{(2\pi)^4}
        \frac{d^4k}{(2\pi)^4}
        G(p)G(q)G(p-k)
         G(q+k)G(k)
        \frac{J(p+q)}{1-\frac{i\lambda^3\phi^2}{4}J(p+q)}
\end{equation}
We replace $p$ by $p-q$ and $k$ by $-k$ in the preceding expression
and assuming $G(k)$ as even function of $k$, obtain
\begin{equation}
V^{(a)}_2(\phi, G)= -\frac{i\lambda^6\phi^4}{96}
\int \frac{d^4p}{(2\pi)^4}
\frac{J^2(p)}{1-\frac{i\lambda^3\phi^2}{4}J(p)}
\label{Gamma2a}
\end{equation}
We shall now discuss about the 2PI diagrams of another kind which
we name it as type-$b$. Consider a diagram of this kind in FIG.\ref{lonli0},
which contains only $n$ lobes without any trilinear line. The contribution
reads as
\begin{eqnarray}
	\gamma_b^{(n,0)} &=& \left(\frac{-i\lambda}{4!}\right)^n
\int \frac{d^4p}{(2\pi)^4} \frac{d^4q}{(2\pi)^4} 
\frac{d^4k_1}{(2\pi)^4}\cdots\frac{d^4k_{n-1}}{(2\pi)^4}
       4G(p)3G(q)4G(k_1)3G(p+q-k_1)\cdots
4G(k_{n-2})\nonumber\\
	& &\times 3G(p+q-k_{n-2})2G(k_{n-1})
        G(p+q-k_{n-1})\nonumber\\
        &=&-\frac{i\lambda}{12}\int\frac{d^4p}{(2\pi)^4} 
        \frac{d^4q}{(2\pi)^4}G(p)G(q)
        \left(\frac{-i\lambda}{2}I(p+q)\right)^{n-1},
\end{eqnarray}
where
\begin{equation}
        I(p)=\int\frac{d^4k}{(2\pi)^4}G(k)G(p-k).
        \label{Ip}
\end{equation}
We now place $r$ trilinear lines in $n-1$ lobes ($1\le r\le n-1$)
of FIG.\ref{lonli0} such that each lobe can accommodate only one
trilinear line. This placement
can be made in $ ^{n-1}\mathrm{C}_r$ ways and
one of the them is shown in FIG.\ref{lonlir}.
The contribution reads as
\begin{eqnarray}
	\gamma_b^{(n,r)} &=& \left(\frac{-i\lambda}{4!}\right)^n
        \left(\frac{-i\lambda\phi}{3!}\right)^{2r}
\int \frac{d^4p}{(2\pi)^4} \frac{d^4q}{(2\pi)^4} 
        4G(p)
        3G(q)(3.3.2.4.3\, J(p+q))^r(4.3\, I(p+q))^{n-2-r}
        2I(p+q)\nonumber\\
        &=& -\frac{i\lambda}{12}\int\frac{d^4p}{(2\pi)^4} 
        \frac{d^4q}{(2\pi)^4}G(p)G(q)
        \left(\frac{i\lambda^3\phi^2}{4}J(p+q)\right)^r
        \left(\frac{-i\lambda}{2}I(p+q)\right)^{n-1-r}
\end{eqnarray}
We take the sum over $r$ ($0\le r\le n-1$) and then over $n$
($2\le n <\infty$) to include the contributions of all possible diagrams
of type-$b$ to the effective potential.
The complete expression is given by
\begin{eqnarray}
        -iV^{(b)}_2(\phi, G)
	&=& \sum_{n=2}^\infty\sum_{r=0}^{n-1}
        { ^{n-1}\mathrm{C}_r}\gamma_b^{(n,r)}
        =-\frac{i\lambda}{12} \int \frac{d^4p}{(2\pi)^4} 
        \frac{d^4q}{(2\pi)^4} G(p)G(q)          
        \sum_{n=2}^\infty 
        \left(\frac{i\lambda^3\phi^2}{4}J(p+q) 
        - \frac{i\lambda}{2}I(p+q)\right)^{n-1}\nonumber\\
        &=& \frac{i\lambda}{12} \int \frac{d^4p}{(2\pi)^4} 
        \frac{d^4q}{(2\pi)^4} G(p)G(q)          
\frac{-\frac{i\lambda^3\phi^2}{4}J(p+q)+\frac{i\lambda}{2}I(p+q)}
        {1-\frac{i\lambda^3\phi^2}{4}J(p+q)+\frac{i\lambda}{2}I(p+q)}.
\end{eqnarray}
We replace $p$ by $p-q$ in the preceding expression and obtain
\begin{equation}
        V^{(b)}_2(\phi, G) =
        -\frac{i\lambda^2}{24} \int \frac{d^4p}{(2\pi)^4} I(p)
        \frac{I(p) -\frac{\lambda^2\phi^2}{2}J(p)}
        {1-\frac{i\lambda^3\phi^2}{4}J(p)+\frac{i\lambda}{2}I(p)}.
\label{Gamma2b}
\end{equation}
The diagram shown in FIG.\ref{g2c} of type say $c$, which is a 
2PI, two loop diagram of order $\lambda$ and is absent in the 
first two types, is considered to contribute to the effective potential.
Its contribution to the potential reads as,
\begin{equation}
V_2^{(c)}(\phi, G)
        = \frac{\lambda}{8}\left(\int\frac{d^4p}{(2\pi)^2} G(p)\right)^2.
\end{equation}
Therefore, the contributions of 2PI diagrams of type-$a$, $b$ and $c$ to the
effective potential are given by
\begin{equation}
V_2(\phi, G)
=V^{(a)}_2(\phi, G)
+V^{(b)}_2(\phi, G)
+V^{(c)}_2(\phi, G)
\end{equation}

\section{Solutions of
$\frac{\partial V_{eff}}{\partial S(k)}=0$ and 
$\frac{\partial V_{eff}}{\partial G(k)}=0$}
\label{delvdelg}

We use the stationary condition Eq.(\ref{Seqn1}) under the variation
of $S(k)$, we obtain the equation as
\begin{equation}
	iS^{-1}(\phi,k) = i{\cal{S}}^{-1}(\phi,k) 
	= k\!\!\!/ - M_f(\phi).
\end{equation}
Then the use of the stationary condition of Eq.(\ref{Geqn1})
in Eq.(\ref{Veff1}) leads to the equation for $G$ as
\begin{equation}
iG^{-1}(k) - iG^{-1}_{ct}(k)
= i{\cal{D}}^{-1}(\phi, k) 
	- 2\sum_{j=a, b, c}\frac{\partial V_2^{(j)}}{\partial G(k)}
\end{equation}
We shall try to solve the equation using the method of iteration.
We choose the trial solution for $G^{-1}(k)$ as ${\cal{D}}^{-1}(k)$
assuming that $M^2(\phi) > k^2$ and $\phi^2$.
Therefore, the terms such as $k^2/M^2(\phi)$ or $\phi^2/M^2(\phi)$
will be ignored for their powers more than one.
Thus the solution reads as,
\begin{equation}
iG^{-1}(k) - iG^{-1}_{ct}(k)
= i{\cal{D}}^{-1}(\phi, k)
	- 2\sum_{j=a, b, c}\frac{\partial V_2^{(j)}}{\partial G(k)}
        \Bigg{|}_{G={\cal{D}}}.
\label{Geqn2}
\end{equation}
The outline for the computations of the
$\frac{\partial V_2^{(j)}}{\partial G(k)}
\Big{|}_{G={\cal{D}}}$ for
$j=a, b, c$ are shown in Appendices A, B and C.
Hence, the solution after one iteration reads as
\begin{eqnarray}
iG^{-1}(k) - iG^{-1}_{ct}(k)
&=& \frac{\lambda}{96\pi^2\epsilon}\Big(m^2+\frac{7}{2}\lambda\phi^2\Big)
        +\Big(1+\frac{\lambda^2\phi^2}{96\pi^2M^2(\phi)}\Big)k^2
- M^2(\phi) + \frac{\lambda M^2(\phi)}{96\pi^2}
        \Big(1-\gamma_E-\ln\Big(\frac{M^2(\phi)}{4\pi\mu^2}\Big)\Big)
\nonumber\\
& & + \frac{\lambda^2 \phi^2}{48\pi^2}
        \Big(2-3\gamma_E-3\ln\Big(\frac{M^2(\phi)}{4\pi\mu^2}\Big)\Big)
 + \frac{f^2(\phi)}{3 M^2(\phi)}
\exp\Big\{b_1\Big(\frac{f(\phi)}{M^2(\phi)}\Big)^2
        -b_2\Big(\frac{f(\phi)}{M^2(\phi)}\Big) +b_3\Big\},
\end{eqnarray}
where
\begin{equation}
f(\phi)=\frac{\lambda^3\phi^2}{1024\pi^4}.
\label{fphi}
\end{equation}
Choose
\begin{equation}
iG^{-1}_{ct}(k) = -
        \frac{\lambda}{96\pi^2\epsilon}\Big(m^2+\frac{7}{2}\lambda\phi^2\Big)
        + \delta m^2,
\label{Ginvct}
\end{equation}
where $\delta m^2$ is finite in the limit $\epsilon\rightarrow 0$.
The renormalized propagator is
\begin{equation}
iG^{-1}(k) = Z k^2 - M_1^2(\phi),
\label{Gp}
\end{equation}
where
\begin{eqnarray}
Z &=& 1+\frac{\lambda^2\phi^2}{96\pi^2M^2(\phi)}\nonumber\\
M_1^2(\phi) &=& 
M^2(\phi) - \delta m^2 
- \frac{\lambda M^2(\phi)}{96\pi^2}
        \Big(1-\gamma_E-\ln\Big(\frac{M^2(\phi)}{4\pi\mu^2}\Big)\Big)
- \frac{\lambda^2 \phi^2}{48\pi^2}
        \Big(2-3\gamma_E-3\ln\Big(\frac{M^2(\phi)}{4\pi\mu^2}\Big)\Big)
\nonumber\\
        & & - \frac{f^2(\phi)}{3 M^2(\phi)}
\exp\Big\{b_1\Big(\frac{f(\phi)}{M^2(\phi)}\Big)^2
        -b_2\Big(\frac{f(\phi)}{M^2(\phi)}\Big)+b_3\Big\}.
\end{eqnarray}
To determine $\delta m^2$, we use the renormalization condition:
$M_1^2(0)=m^2$ and obtain
\begin{equation}
\delta m^2 = -\frac{\lambda m^2}{96\pi^2}
\Big(1-\gamma_E -\ln\Big(\frac{m^2}{4\pi\mu^2}\Big)\Big).
\end{equation}
Therefore,
\begin{eqnarray}
	M_1^2(\phi) &=& M^2(\phi) 
+ \frac{\lambda m^2}{96\pi^2}\ln\Big(\frac{M^2(\phi)}{m^2}\Big)
- \frac{\lambda^2 \phi^2}{192\pi^2}
        \Big(9-13\gamma_E-13\ln\Big(\frac{M^2(\phi)}{4\pi\mu^2}\Big)\Big)
	\nonumber\\
	& &- \frac{f^2(\phi)}{3 M^2(\phi)}
\exp\Big\{b_1\Big(\frac{f(\phi)}{M^2(\phi)}\Big)^2
        -b_2\Big(\frac{f(\phi)}{M^2(\phi)}\Big)+b_3\Big\}.
\end{eqnarray}
\begin{figure}[!ht]
\begin{center}
\includegraphics[scale=0.30]{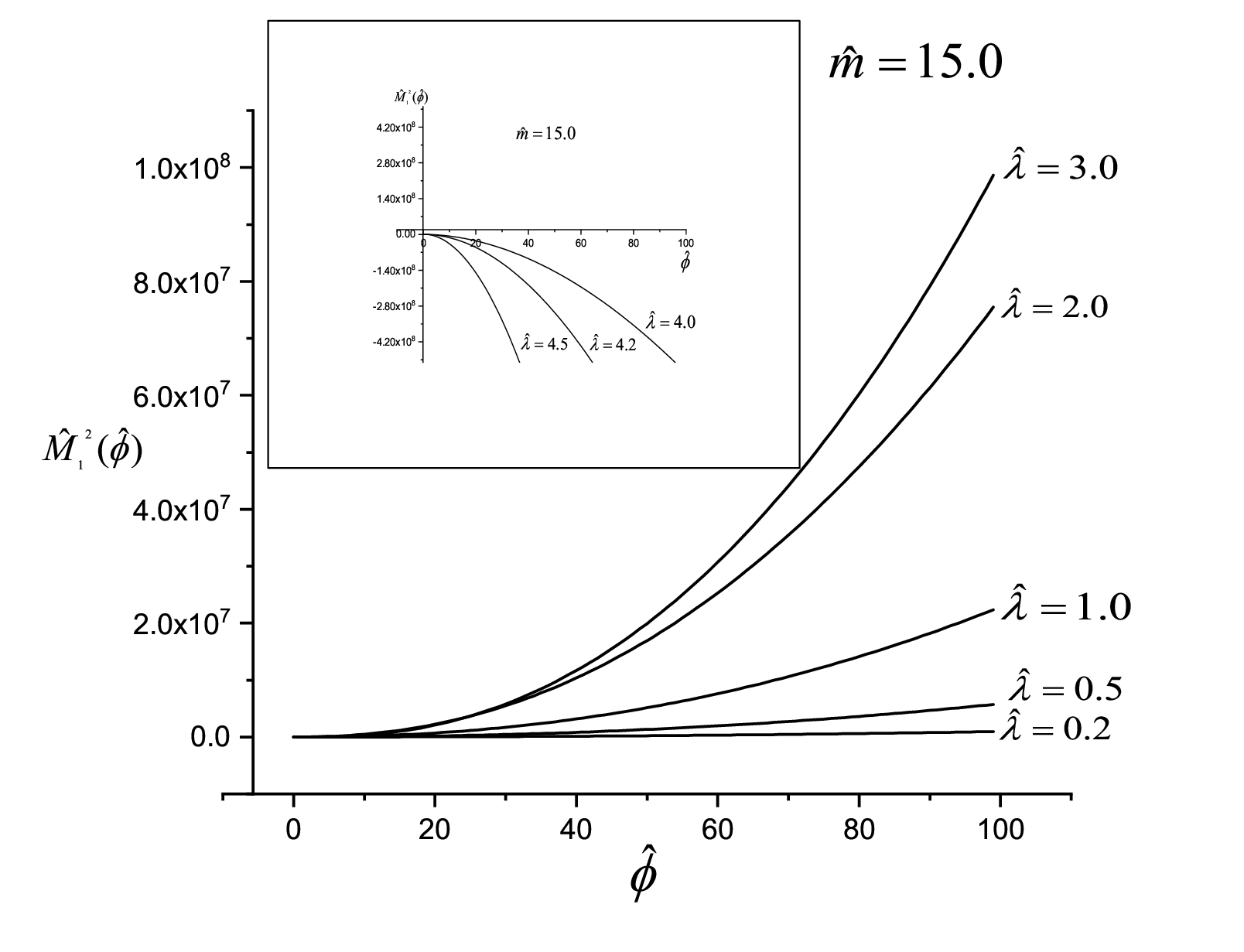}
\end{center}
\vskip -0.37in
        \caption[]{Plots of $\hat{M_1^2}(\hat{\phi})$ versus $\hat{\phi}$
        for different values of $\hat{\lambda}$.
        Here, $\hat{\lambda}
        =\lambda/16\pi^2$ and $\hat{\phi}=\phi/\sqrt{4\pi\mu^2}$.}
        \label{gap1}
\end{figure}
To check the validity of the solution we plot $M_1^2(\phi)$ as
a function of $\phi$ for different $\lambda$. The plot is
shown in Fig. \ref{gap1}.
$M_1^2$ increases rapidly with $\phi$ for $\hat{\lambda}\le 3.0$.
When $\hat{\lambda}> 3.0$, $M_1^2$ attains negative value for 
$\hat{\phi}\ge 0$ (shown in the inset plot).
The indications are of two-fold: On the one hand it suggests that
more than one iterations may be required to make 
$M_1^2 >0$ in the region of $\hat{\lambda}$ beyond $3.0$. 
On the other hand it may be assumed that  
apart from those of $a$, $b$ and $c$ types already taken into considerations,  
the contributions of the other types of diagrams, 
are to be included to get a positive result
for $M_1^2(\phi)$ obtained after first iteration for $\hat{\lambda} > 3.0$.

\section{Effective Potential}
\label{effectivepotential}

The classical potential reads as
\begin{equation}
        V(\phi)=\frac{1}{2}m^2\phi^2
        +\frac{\lambda}{4!}\phi^4.
\end{equation}
The remaining parts of Eq.(\ref{Veff1}) describe the quantum corrections
over the classical part of the potential.
The integrals in $V_{eff}$ are evaluated using Eq.(\ref{Gp}), the
details of which are shown in the Appendix-D.
The renormalized effective potential is obtained as
\begin{equation}
        V_{eff}(\phi) = V(\phi)+V_{qc}(\phi)+V_d(\phi)+V_{ct}(\phi),
\end{equation}
where the quantum corrections and its ultraviolet divergent
contributions are contained in $V_{qc}$ and $V_d$ respectively.
The expressions for them are as follows:
\begin{eqnarray}
	V_d(\phi) &=& \sum_{j=1}^4I_{j(div)}
	+\sum_{j=a,b,c}V_{2(div)}^{(j)}(\phi, G)\nonumber\\ 
&=& -\frac{\lambda M_1^2(\phi)}{6144\pi^4Z^4\epsilon^2}
\Big[Z^2\Big(m^2+\frac{7}{2}\lambda\phi^2\Big)-M_1^2(\phi)\Big]
-\frac{a_1^2}{96\pi^2\epsilon}\Big(\frac{\lambda^3\phi^2}
{1024\pi^4Z^5}\Big)^2
+\frac{M_1^4(\phi)}{32\pi^2Z^2\epsilon}\Big(\frac{1}{Z}
-\frac{1}{2}+\frac{M^2(\phi)}{M_1^2(\phi)}\Big)\nonumber\\
&-&\frac{\lambda M_1^2(\phi)}{3072\pi^4Z^2\epsilon}
\Big[m^2\Big(1-\gamma_E-\ln\Big(\frac{m^2}{4\pi\mu^2}\Big)\Big)
+\Big(m^2+\frac{7}{2}\lambda\phi^2-M_1^2(\phi)\Big)
\times\Big(1-\gamma_E
-\ln\Big(\frac{M_1^2(\phi)}{4\pi\mu^2Z}\Big)\Big)\Big]\nonumber\\
	& & +\frac{M_f^4(\phi)}{32\pi^2\epsilon},
\end{eqnarray}
\begin{eqnarray}
	V_{qc}(\phi) &=& \sum_{j=1}^4I_{j(fin)}
	+\sum_{j=a,b,c}V_{2(fin)}^{(j)}(\phi, G)\nonumber\\ 
&=& - \frac{M_1^4(\phi)}{128\pi^2 Z^2}
        +\frac{a_1a_2\lambda^2\phi^2M_1^2(\phi)}{3072\pi^4Z^4}
        -\frac{M_1^2(\phi)}{64\pi^2Z^2}\Big(M_1^2(\phi)
        -\frac{2M_1^2(\phi)}{Z}+M^2(\phi)\Big)\Big(1-\gamma_E
	-\ln\Big(\frac{M_1^2(\phi)}{4\pi\mu^2Z}\Big)\Big)\nonumber\\
        &-&\frac{\lambda M_1^2(\phi)}{6144\pi^4Z^2}
	\Big(m^2+\frac{7}{2}\lambda\phi^2\Big)
        \Big[1+\frac{\pi^2}{6}+\Big(1-\gamma_E-
	\ln\Big(\frac{M_1^2(\phi)}{4\pi\mu^2Z}\Big)\Big)^2\Big]
        -\frac{\lambda M_1^2(\phi)m^2}{3072\pi^4Z^2}
	\Big(1-\gamma_E-\ln\Big(\frac{m^2}{4\pi\mu^2}\Big)\Big)\nonumber\\
        &\times&\Big(1-\gamma_E
	-\ln\Big(\frac{M_1^2(\phi)}{4\pi\mu^2Z}\Big)\Big)
        +\frac{\lambda M_1^4(\phi)}{6144\pi^4Z^2}
        \Big[1+\frac{\pi^2}{6}+2\Big(1-\gamma_E
	-\ln\Big(\frac{M_1^2(\phi)}{4\pi\mu^2Z}\Big)\Big)^2\Big]\nonumber\\
&-&\frac{a_1f(\phi)}{96\pi^2Z^2}
        \Big[M_1^2(\phi)a_2\ln{a_2} + a_1(1-\gamma_E)f(\phi)
        - (a_2M_1^2(\phi)+a_1f(\phi))\ln\Big(\frac{a_2M_1^2(\phi)+a_1f(\phi)}
{4\pi\mu^2Z}\Big)\nonumber\\
        &-& \frac{2a_3f(\phi)}{\Delta^2(\phi)}
        \{(a_4-a_2)^2M_1^4(\phi) - a_1(2a_4-3a_2)M_1^2(\phi)f(\phi)
+ 2a_1^2f^2(\phi)\}\nonumber\\
        &+& \frac{1}{\Delta(\phi)}
        \{a_2(a_4-a_2)M_1^4(\phi) - a_1(4a_4-3a_2)M_1^2(\phi)f(\phi)
        + 4a_1^2f^2(\phi)\}\ln\Big(\frac{M_1^2(\phi)}{4\pi\mu^2Z}\Big)\Big]
\nonumber\\
&-&\frac{M_f^4(\phi)}{32\pi^2}\Big[\frac{1}{2}+\gamma_E
+ \ln\Big(\frac{M_f^2(\phi)}{4\pi\mu^2}\Big)\Big],
\end{eqnarray}
where $\Delta(\phi) = (a_4-a_2)M_1^2(\phi)- a_1f(\phi)$. 
Here, $I_{j(fin)}$ and $I_{j(div)}$ ($1\le j\le 4$) denote 
finite and divergent parts of 
Eq.(\ref{I1}), (\ref{I2}), (\ref{I3}), (\ref{I4}) respectively
in the $\epsilon\rightarrow 0$ limit.
Similarly, $V_{2(fin)}^{(k)}$ and $V_{2(div)}^{(k)}$ ($k=a, b, c$)
denote finite and divergent parts of Eq.(\ref{Gamma2a2}), (\ref{Gamma2b2}) 
and (\ref{Gamma2c2}) respectively. 
We expand $V_{d}(\phi)$and $V_{ct}(\phi)$ in powers of $\phi^2$:
\begin{equation}
	V_{d}(\phi)=\sum_{j=0}^{\infty}v_{d}^{(j)}\phi^{2j},
\end{equation}

\begin{equation}
	V_{ct}(\phi)=\sum_{j=0}^{\infty}(v_{ct}^{(j)}
	+\delta v_{ct}^{(j)})\phi^{2j},
\end{equation}
where $\delta v_{ct}^{(j)}$s are the finite parts of the counter terms 
that are determined using renormalization conditions. Thus we obtain
\begin{equation}
	V_{eff}(\phi)=V(\phi)+V_{qc}(\phi)+\sum_{j=0}^{\infty}(v_{d}^{(j)}
+v_{ct}^{(j)}+\delta v_{ct}^{(j)})\phi^{2j}.
\end{equation}
To remove the infinities present in the $V_{eff}$, we choose the 
counter terms as
\begin{equation}
	v_{ct}^{(j)}=-v_{d}^{(j)},
\end{equation}
for $0\le j \le \infty$. We use the following renormalization conditions 
to determine $\delta v_{ct}^{(j)}$ ($0\le j \le \infty$):
\begin{equation}
	V_{eff}(0)=0,
\end{equation}
\begin{equation}
	\frac{d^2V_{eff}}{d\phi^2}\Big{|}_{\phi=0}=m^2,
\end{equation}
\begin{equation}
	\frac{d^4V_{eff}}{d\phi^4}\Big{|}_{\phi=s\sqrt{4\pi\mu^2}}=\lambda,
       (0<s\le 0.1)
\label{fourth derivative}
\end{equation}
\begin{equation}
	\delta v_{ct}^{(j)}=0  (j \ge 3).
\end{equation}
The part of $V_{qc}$ coming from fermionic contribution is proportional
to $\phi^4\ln(\phi^2/4\pi\mu^2)$ which is divergent when the fourth
derivative of the term is evaluated at $\phi=0$. To avoid this problem, the 
fourth derivative is evaluated at $\phi =s\sqrt{4\pi\mu^2}$ in 
Eq.(\ref{fourth derivative}).
The use of these conditions led to the following expressions 
for $\delta v_{ct}^{(j)}$ for $0\le j \le 2$:
\begin{equation}
	\delta v_{ct}^{(0)} = -V_{qc}(0) = \frac{m^4}{128\pi^2}
        +\frac{\lambda m^4}{6144\pi^4}
	\Big(1-\gamma_E-\ln\Big(\frac{m^2}{4\pi\mu^2}\Big)\Big)^2.
\end{equation}
\begin{equation}
	\delta v_{ct}^{(1)}
	=-\frac{1}{2}\frac{d^2V_{qc}}{d\phi^2}\Big{|}_{\phi=0}.
\end{equation}
\begin{equation}
        \delta v_{ct}^{(2)}
	=-\frac{1}{4!}\frac{d^4V_{qc}}{d\phi^4}\Big{|}_{\phi=s\sqrt{4\pi\mu^2}}.
\end{equation}
Hence, the renormalized effective potential takes the form
\begin{equation}
	V_{eff}(\phi)=V(\phi)+V_{qc}(\phi)+\delta v_{ct}^{(0)}+\delta v_{ct}^{(1)}\phi^2+\delta v_{ct}^{(2)}\phi^4.
\end{equation}
\begin{figure}[!ht]
\begin{center}
\includegraphics[scale=0.30]{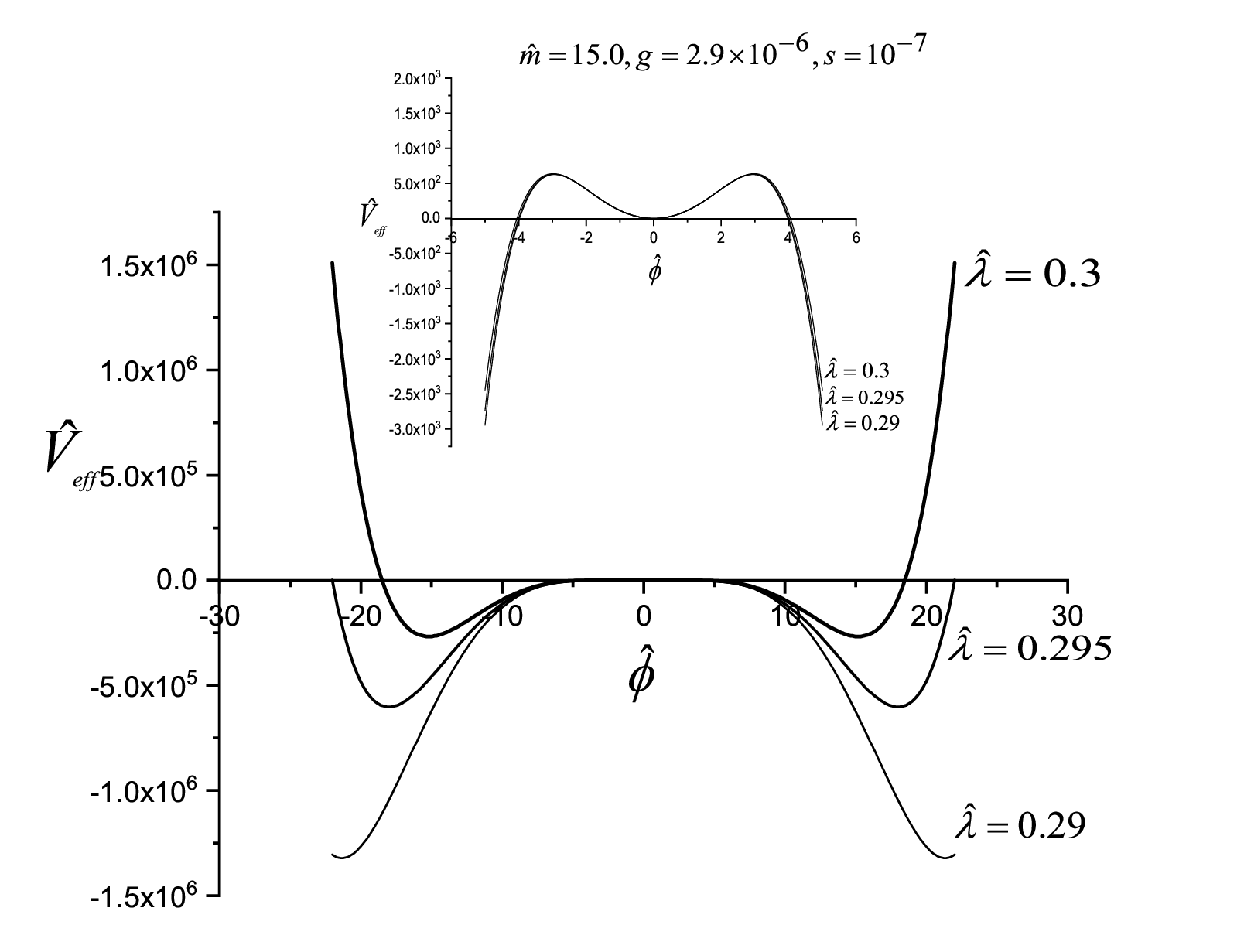}
\end{center}
\vskip -0.37in
\caption[]{Plots of $\hat{V}_{eff}$ versus $\hat{\phi}$
        for $\hat{m}=15.0$ and for different values of $\hat{\lambda}$.}
	\label{pot1}
\end{figure}

\begin{figure}[!ht]
\begin{center}
\includegraphics[scale=0.30]{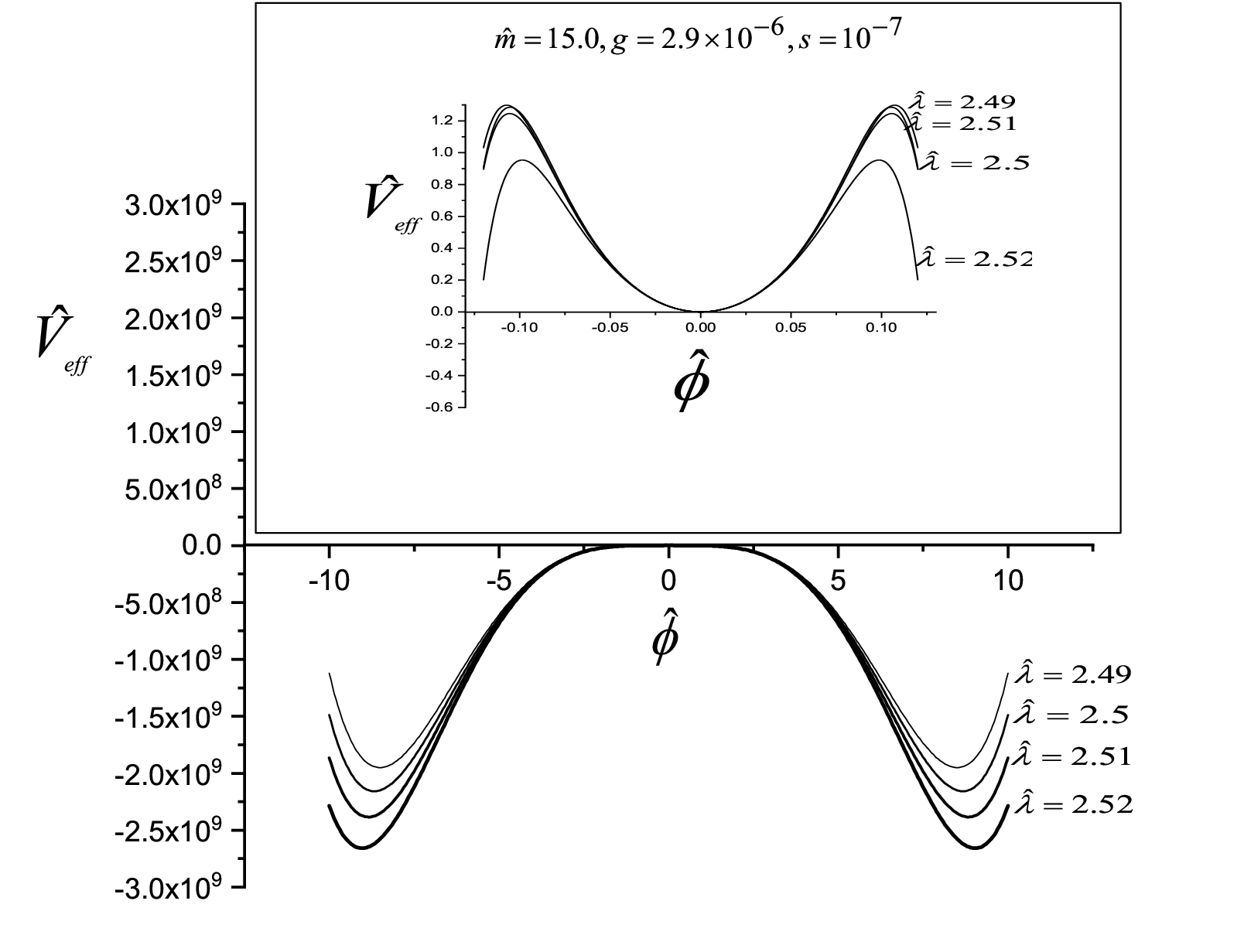}
\end{center}
\vskip -0.37in
\caption[]{Plots of $\hat{V}_{eff}$ versus $\hat{\phi}$
        for $\hat{m}=15.0$ and for different values of $\hat{\lambda}$.}
\label{pot2}
\end{figure}
In Fig. \ref{pot1} and Fig. \ref{pot2}, it is observed that the potential 
exhibits maxima and minima above and below the zero potential line respectively at non-zero $\phi$ on either side of the minimum at $\phi=0$.
The potential has two non-trivial minima below
the $V_{eff}=0$ line at non-zero $\phi$, that are symmetrically
situated on either side of the minimum at $\phi=0$.
This is due to the presence of symmetry under $\phi\rightarrow -\phi$
in the effective potential.

It is evident from the data given in the Table \ref{table1} 
that minima of the potential shift towards $\phi=0$
with the increase of $\hat{\lambda}$ for
$\hat{\lambda} \le 0.32$ and shift away from the point $\phi=0$ with 
increase of $\hat{\lambda}$ for $1.6 \le \hat{\lambda} \le 3.0$.
The depth of the minima decrease with increase of
$\hat{\lambda}$ for $\hat{\lambda} \le 0.32$ and it increases with 
increase of $\hat{\lambda}$ for $1.6 \le \hat{\lambda} \le 3.0$.
Moreover, the positions of maxima shift towards $\phi=0$
and its height decrease with 
increase of $\hat{\lambda}$ for $0\le\hat{\lambda}\le 0.32$ 
and $1.6 \le \hat{\lambda} \le 3.0$. 
It is also noted from the results of our numerical evaluations
that the effective potential exhibits only a minimum at $\phi=0$ for 
$0.32<\hat{\lambda}<1.6$.
\begin{table}
	\caption{\label{table1}$d$ and $h$ versus $\hat{\lambda}$, where 
	$d= |\hat{V}_{eff}(\pm\phi_{min})|$ = depth of the potential at the
	minimum $\pm \phi_{min}$ and $h= |\hat{V}_{eff}(\pm\phi_{max})|$
	= height of the potential at the maximum $\pm \phi_{max}$.}
\centering
\begin{tabular}{l c c c c }
\hline
$\hat{\lambda}$ & $\pm\phi_{max}$ & $h$ & $\pm\phi_{min}$ & $d$\\
\hline
$0.1$ & $9.39$ & $7423.49$ & $5.4\times 10^{7}$ & $3.3\times 10^{31}$\\
$0.2$ & $3.78$ & $1084.02$ & $953.98$ & $8.3\times 10^{12}$\\
$0.29$ & $2.94$ & $627.48$  & $21.29$ & $1.3\times 10^{6}$\\
$0.295$ & $2.96$ & $629.81$ & $18.04$ & $6.03\times 10^{5}$\\
$0.3$ & $2.98$ & $636.71$ & $15.16$ & $2.7\times 10^{5}$\\
$0.31$ & $3.09$ & $670.36$ & $10.72$ & $4.5\times 10^{4}$\\
$0.32$ & $3.35$ & $763.33$ & $7.28$ & $3.4\times 10^{3}$\\
$1.6$ & $0.67$ & $974.74$ & $1.66$ & $4.48\times 10^{4}$\\
$1.8$ & $0.40$ & $153.71$ & $2.60$ & $1.67\times 10^{6}$\\
$2.0$ & $0.27$ & $36.06$ & $3.65$ & $1.6\times 10^{7}$\\
$2.2$ & $0.18$ & $8.06$ & $5.08$ &  $1.2\times 10^{8}$\\
$2.4$ & $0.14$ & $4.05$ & $7.15$ & $7.8\times 10^{8}$\\
$2.49$ & $0.11$ & $1.30$ & $8.50$ & $1.95\times 10^{9}$\\
$2.5$ & $0.11$ & $1.25$ & $8.67$ & $2.2\times 10^{9}$\\
$2.51$ & $0.11$ & $1.29$ & $8.85$ & $2.4\times 10^{9}$\\
$2.52$ & $0.098$ & $0.95$ & $9.05$ & $2.7\times 10^{9}$\\
$2.8$ & $0.066$ & $0.38$ & $17.05$ & $6.2\times 10^{10}$\\
$3.0$ & $0.055$ & $0.28$ & $31.20$ & $1.0\times 10^{12}$\\
\hline
\end{tabular}
\label{tab1}
\end{table}
It is also observed from our numerical studies that 
the positions of maxima and minima and their heights and depths 
are respectively remain unchanged when $g$ lies in the range  
$10^{-8}\le g\le 1.0$ for any $\hat{\lambda}$ lying in the 
range $0\le\hat{\lambda}\le 0.32$ and $1.6 \le \hat{\lambda} \le 3.0$.
It indicates that the fermionic part plays a less dominant role 
than the scalar part of the potential in ascertaining the extremum
of the potential.

\section{Discussions}
\label{results and discussions}

In this paper we have taken a scalar theory with 
$\lambda\phi^4$ interaction, coupled to a massless fermion through Yukawa 
interaction. We have tried to find an answer to a query whether 
the inclusion of infinite loop corrections in a spontaneously 
unbroken theory described by $m^2>0$ and $\lambda > 0$ may lead
to an appearance of minima occurring at non-zero $\phi$ so that 
the fermion can acquire mass in the new vacuum through Yukawa 
interaction.  
We have used CJT method which provides a means to obtain  
the effective potential as a functional of the scalar and 
fermion correlation functions $G$ and $S$ respectively for 
constant $\phi$.  
The effective potential is obtained 
as sum of the classical potential, the one loop corrections
and the contributions of two and higher loop, 2PI diagrams
that are denoted collectively as $V_2(\phi, G, S)$.
We have included quantum corrections of infinite loop,
2PI diagrams of two different types, namely $a$ and $b$ and of
a single two loop, 2PI diagram of type $c$ to $V_2(\phi, G, S)$.
Since, the infinite sum of diagrams of type $a$ and $b$ are obtained 
in closed forms which do not make reference to the smallness of the 
coupling constant, the effective potential so computed 
is non-perturbative in nature.
The requirements that it remains stationary under the independent 
variations of $G$ and $S$ give two equations. 
The solution of one of the equations give $S$ as a Dirac 
propagator of mass equals to $g\phi$. 
The equation for $G$ involves ultraviolet divergence which is 
removed choosing counterterm in the effective potential and 
its finite part is obtained employing the renormalization condition
that the inverse of $G$ at $p^2=m^2$ and $\phi=0$ is zero.  
Then, the equation is solved using the method of iteration retaining terms 
to order $\phi^2/M^2(\phi)$ or $k^2/M^2(\phi)$, where $k$ is the 
external momentum of a Feynman Diagram.
Our numerical evaluation shows that pole of $G$ becomes imaginary 
for very large coupling ($\hat{\lambda} > 3.0$), which may be 
an indications of either the invalidity of the result of 
iteration to first order only or, the requirement of inclusion 
of more diagrams of different types that are not considered in 
our calculations. The $G$ and $S$ obtained as solutions of 
the two equations are used to compute the effective potential 
as a function of $\phi$, which is valid in our case 
for $\hat{\lambda}\le 3.0$.
Since, the ultraviolet divergent terms in the effective potential 
involve non-polynomial functions of $\phi$ through 
$\ln{M_1^2(\phi)/4\pi\mu^2}$, and integer powers of $Z^{-1}(\phi)$, 
we have expanded the terms in powers of $\phi^2$ about $\phi=0$ 
and have introduced infinite number of counter terms 
expressed in power series of $\phi^2$ to remove the divergences. 
The finite parts of first three counter terms are determined 
employing renormalization conditions required to fix the zero
of the potential at $\phi=0$, square of the mass parameter ($m^2$)
and the coupling constant ($\lambda$). The finite parts
of the remaining counter terms are set to zero.
It is quite evident that there are key advantages in using CJT 
method for computation of effective potential for this theory. 
Firstly, it provides the effective, non-perturbative 
propagators which give information about the mass of 
the fields in the new vacuum and helps to find the upper bound of the 
coupling constant for which the pole of the propagator remains real.
Secondly, we have been able to compute a fraction of the total
non-perturbative corrections contributing to the effective potential
using the constraint of two particle irreducibility
of diagrams having more than one loop.

It is clear that owing to non-perturbative corrections to the effective
potential, it develops maxima of small height and minima of very large depth
above and below the zero potential line respectively on either side
of $\phi=0$ for $0 < \hat{\lambda}\le 0.32$ and 
$1.6\le\hat{\lambda}\le 3.0$. 
Moreover, as the coupling constant increases the positions of the maxima
shift rapidly towards $\phi=0$.
On the other hand, as the coupling constant increases 
minima shift towards and away from $\phi=0$ when the 
values are assumed form the regions 
$0 < \hat{\lambda}\le 0.32$
and $1.6\le\hat{\lambda}\le 3.0$ respectively.
For coupling constant lying in the region $0.32 <\hat{\lambda} < 1.6$
the effective potential exhibits only a minimum 
at $\phi=0$. These indicate that 
the quantum corrections dominate over the classical behaviour of the
potential when the coupling constant lies 
in the regions $0 < \hat{\lambda}\le 0.32$ and 
$1.6\le\hat{\lambda}\le 3.0$ and thus the non-perturbative 
quantum corrections lead to form 
maxima and minima of the potential at non-zero $\phi$
in those two regions of the coupling constant. 
However, when the coupling constant lies in the region 
$0.32 <\hat{\lambda} < 1.6$, the classical terms of the 
potential dominate over the quantum corrections and 
ensure the minimum of the potential to stay only at $\phi=0$.
There are numerical evidences suggesting that the scalar part 
plays a dominant role over the fermionic part of the potential 
in ascertaining its extremum.

The effective potential has an inversion symmetry under the transformation  
$\phi\rightarrow -\phi$. Since the system always settles into  
a lowest minimum of the potential, the fermion receives mass in a 
non-trivial minima at positive $\phi$ when the value of the 
coupling constant falls in any one of the following regions:
$0 < \hat{\lambda}\le 0.32$ and $1.6\le\hat{\lambda}\le 3.0$.
This breaks the inversion symmetry of the vacuum because
one of the minima occurring at non-zero $\phi$ is chosen as the 
vacuum of the theory.
Even if the system is somehow be trapped into the minimum at $\phi=0$, 
later it may fall into the lowest minimum at non-zero $\phi$ acquiring 
small amount of kinetic energy at relatively large coupling constant 
because the barrier height rapidly diminishes with the increase of 
coupling constant. Hence, the minimum at $\phi=0$ becomes unstable
at large coupling constant. 
However, when the value of the coupling constant falls in the region 
$0.32 <\hat{\lambda} < 1.6$, the system stays in the minimum
at $\phi=0$ and the fermion remains massless. 

\appendix

\section{Evaluations of $I(p)$ and $J(p)$ for $G={\cal{D}}$}
\label{IpJp}

The integral in Eq.(\ref{Ip}) gives
\begin{equation}
        I(p) = \int\frac{d^4k}{(2\pi)^4}{\cal{D}}(k)
{\cal{D}}(p-k).
\end{equation}
The ultraviolet divergent integral has been regularized by going
over to $d$ dimension, where $d=4-2\epsilon$ ($\epsilon >0$) and the
result\cite{Collins} obtained after performing the integration is
\begin{equation}
        I(p)=-\frac{i\Gamma(1+\epsilon)}{16\pi^2\epsilon}
        \left(\frac{M^2(\phi)}{4\pi\mu^2}\right)^{-\epsilon}\hat{I}(p),
        \label{Ip1}
\end{equation}
where
\begin{equation}
        \hat{I}(p)=\int_0^1 dx 
        \left[1-\frac{p^2}{M^2(\phi)}x(1-x)\right]^{-\epsilon}
\label{Ihatp}
\end{equation}
and $\mu$ is the scale of mass dimension $1$ associated with
the theory.

The integral in Eq.(\ref{Jp}) gives
\begin{equation}
        J(p) = \int\frac{d^4q}{(2\pi)^4}\int\frac{d^4r}{(2\pi)^4}
        {\cal{D}}(q){\cal{D}}(p-q){\cal{D}}(q-r)
        {\cal{D}}(p-q+r){\cal{D}}(r)
\end{equation}
Though the integral is ultraviolet convergent, it has been evaluated
in $d$ dimension. We go over to the Euclidean space\cite{Raymond} so that
the Euclidean $4$ momentum $p_E=(p_4, {\bf{p}})$, where $p_4=ip_0$ and
$p_E^2=p_4^2+{\bf{p}}^2$. The integral becomes
\begin{eqnarray}
	& &J(p_E)
	=\frac{i\mu^{8-2d}}{(2\pi)^{2d}}\nonumber\\
	& &\times \int 
	\frac{ d^dq_E d^dr_E}
	{(q_E^2+M^2(\phi))
        ((p_E-q_E)^2+M^2(\phi)) ((q_E-r_E)^2+M^2(\phi))
        ((p_E-q_E+r_E)^2+M^2(\phi))
        (r_E^2+M^2(\phi))}.
\end{eqnarray}
We use the Schwinger's parametrization\cite{Collins} and write it as
\begin{eqnarray}
	J(p_E) &=& \frac{i\mu^{8-2d}}{(2\pi)^{2d}}\int 
 d^dq_E d^dr_E
\int_0^\infty d\alpha_1\cdots d\alpha_5
        \exp[-\alpha_1(q_E^2+M^2)
        -\alpha_2((p_E-q_E)^2+M^2)\nonumber\\
        & &-\alpha_3((q_E-r_E)^2+M^2) 
        - \alpha_4((p_E-q_E+r_E)^2+M^2)
        -\alpha_5(r_E^2+M^2)].
\end{eqnarray}
Performing the integrations over $q_E$ and $r_E$ and rescaling
$\alpha_i$ as $\alpha_iM^2(\phi)=x_i$ ($1\le i\le 5$) we obtain
\begin{equation}
        J(x) = \frac{i}{256\pi^4 M^2(\phi)}
        \Big[\Big\{1-2\epsilon\ln\Big(\frac{M^2(\phi)}{4\pi\mu^2}
        \Big)\Big\}J_1(x)
+ \epsilon J_2(x)+ 0(\epsilon^2)\Big],
\label{Jx}
\end{equation}
where,
\begin{eqnarray}
        x &=& p_E^2/M^2(\phi),\\
J_1(x) &=& \int_0^\infty\frac{dx_1\cdots dx_5}{\eta_1^2}
        \exp[-\eta_2x-(x_1+\cdots+x_5)],\\
J_2(x) &=& \int_0^\infty\frac{dx_1\cdots dx_5}{\eta_1^2}
        \exp[-\eta_2x-(x_1+\cdots+x_5)]
        \ln{\eta_1},\\
        \eta_1 &=& (x_1+x_2+x_5)(x_3+x_4)+x_5(x_1+x_2),\\
\eta_2 &=& (x_1+\cdots+x_4)^{-1}\Big[(x_1+x_3)(x_2+x_4)
- \frac{(x_1x_4-x_2x_3)^2}{(x_1+x_2+x_5)(x_3+x_4)+x_5(x_1+x_2)}
\Big].
\end{eqnarray}
Our numerical evaluation suggests that $J_1$ and $J_2$ will be
of following forms:
\begin{eqnarray}
        J_1(x) &=& \frac{a_1}{x+a_2}, \label{J1x}\\
        J_2(x) &=& -\frac{a_3}{x+a_4},\label{J2x}
\end{eqnarray}
where $a_1=5.91\pm 0.05$, $a_2=7.69\pm 0.09$, $a_3=51.54\pm 0.83$
and $a_4=23.55\pm 0.48$.

\section{Evaluation of
$\frac{\partial V_{2}^{(a)}}{\partial G(k)}
\Big{|}_{G={\cal{D}}}$}
\label{dGamma2a}

\begin{equation}
\frac{\partial V_{2}^{(a)}}{\partial G(k)}
\Big{|}_{G={\cal{D}}}
        = - \sum_{i=1}^5 T^{(a)}_i(k),
\end{equation}
where
\begin{eqnarray}
        T^{(a)}_1(k) &=& \frac{i\lambda^6\phi^4}{48}
\int\frac{d^4p}{(2\pi)^4}\frac{d^4q}{(2\pi)^4} 
        {\cal{D}}(p-k){\cal{D}}(q){\cal{D}}(k-q)
        {\cal{D}}(p+q-k)
\frac{J(p)}{1-\frac{i\lambda^3\phi^2}{4}J(p)},\\
        T^{(a)}_2(k) &=& \frac{i\lambda^6\phi^4}{96}
\int\frac{d^4p}{(2\pi)^4}\frac{d^4q}{(2\pi)^4} 
        {\cal{D}}(p+q){\cal{D}}(q){\cal{D}}(k-q)
        {\cal{D}}(p+q-k)
\frac{J(p)}{1-\frac{i\lambda^3\phi^2}{4}J(p)},\\
        T^{(a)}_3(k) &=& \frac{i\lambda^6\phi^4}{48}
\int\frac{d^4p}{(2\pi)^4}\frac{d^4q}{(2\pi)^4} 
        {\cal{D}}(p+q){\cal{D}}(q)
        {\cal{D}}(p+q-k){\cal{D}}(k-p)
\frac{J(p)}{1-\frac{i\lambda^3\phi^2}{4}J(p)},\\
        T^{(a)}_4(k) &=&\frac{i\lambda^6\phi^4}{24} \int \frac{d^4p}{(2\pi)^4}
        \frac{d^4q}{(2\pi)^4}\frac{d^4r}{(2\pi)^4}
{\cal{D}}(p+q){\cal{D}}(q+r)
        \frac{{\cal{D}}(p+q+r){\cal{D}}(r){\cal{D}}(q)
        {\cal{D}}(p-k)L_{1}(p,k)}
{(1-\frac{i\lambda^3\phi^2}{4}J(p))^2},\\
        T^{(a)}_5(k) &=&\frac{i\lambda^6\phi^4}{96} \int \frac{d^4p}{(2\pi)^4}
        \frac{d^4q}{(2\pi)^4}\frac{d^4r}{(2\pi)^4}
        {\cal{D}}(p+q){\cal{D}}(q+r)
        \frac{{\cal{D}}(p+q+r){\cal{D}}(r){\cal{D}}(q)L_{2}(p,k)}
{(1-\frac{i\lambda^3\phi^2}{4}J(p))^2}.
\end{eqnarray}
The functions $L_1$ and $L_2$ are given as
\begin{eqnarray}
L_1(p,k) &=& \int \frac{d^4k^{\prime}}{(2\pi)^4} 
        {\cal{D}}(k^{\prime}){\cal{D}}(k^{\prime}+k)
{\cal{D}}(p-k+k^{\prime})\\
L_2(p,k) &=& \int \frac{d^4k^{\prime}}{(2\pi)^4} 
        {\cal{D}}(k^{\prime}){\cal{D}}(k^{\prime}+k)
{\cal{D}}(k^{\prime}+p)
        {\cal{D}}(p+k+k^{\prime}).
\end{eqnarray}
We shall evaluate $T^{(a)}_i$s for $k=0$ because the non-leading
contributions for $k\ne 0$ will be suppressed by powers
of $1/M^2(\phi)$, where $m$ will be taken to be of order of the mass scale $\mu$
associated with the theory. Since, the integrals are convergent
in the ultraviolet limit, those will be evaluated in $4$ dimension.
Consider the integral $T^{(a)}_1$ evaluated at $k=0$:
\begin{equation}
        T^{(a)}_1(0) = \frac{i\lambda^6\phi^4}{48}
\int\frac{d^4p}{(2\pi)^4}\frac{d^4q}{(2\pi)^4} 
{\cal{D}}(p){\cal{D}}^2(q){\cal{D}}(p+q)
 \frac{J(p)}{1-\frac{i\lambda^3\phi^2}{4}J(p)}.
\end{equation}
Now we go over to the Euclidean space and after parameterizing
$q_E$ dependent denominators using the method
of Schwinger we perform
integration over $q_E$ and obtain
\begin{equation}
T^{(a)}_1(0) = -\frac{i\lambda^6\phi^4}{768\pi^2}
\int_0^\infty \frac{d\alpha_1 d\alpha_2 \alpha_1}
{(\alpha_1+\alpha_2)^2}
\int\frac{d^4p_E}{(2\pi)^4}
        \frac{1}{(p_E^2+M^2(\phi))}
\frac{J(p_E)}{1-\frac{i\lambda^3\phi^2}{4}J(p_E)}
        \exp\Big[-\frac{\alpha_1\alpha_2}{\alpha_1+\alpha_2}p_E^2
        -(\alpha_1+\alpha_2)\Big].
\end{equation}
Eq. (\ref{Jx}) and (\ref{J1x}) give
\begin{equation}
        J(p_E) = \frac{ia_1}{256\pi^4 (p_E^2+a_2M^2(\phi))} + 0(\epsilon).
\end{equation}
Now after scaling the variables $\alpha_i$ as $\alpha_i M^2(\phi) = x_i$
($i=1,2$) and writing $d^4p_E=2\pi^2p_E^3dp_E = \pi^2 M^4(\phi) x dx$
where $x=p_E^2/M^2(\phi)$ we obtain
\begin{equation}
        T^{(a)}_1(0) 
        =\frac{a_1M^2(\phi){b^\prime}^2}{3}
\int_0^\infty dx dx_1 dx_2 \frac{x x_1}{(x+a_2+a_1b^\prime)(x+1)}
\frac{1}{(x_1+x_2)^2}\exp\left[-\frac{x_1x_2x}{x_1+x_2}
-(x_1+x_2)\right],
\end{equation}
where
\begin{equation}
        b^\prime = \frac{\lambda^3\phi^2}{1024\pi^4 M^2(\phi)}
        = \frac{\lambda^3\phi^2}{1024\pi^4 M^2(\phi)}.
\end{equation}
Proceeding with the similar method we obtain following results
for $T^{(a)}_2$ and $T^{(a)}_3$:
\begin{eqnarray}
      T^{(a)}_2(0)
        &=& \frac{a_1M^2(\phi){b^\prime}^2}{6}
\int_0^\infty dx dx_1 dx_2 \frac{x x_1 x_2}{(x+a_2+a_1b^\prime)}
\frac{1}{(x_1+x_2)^2}\exp\Big[-\frac{x_1x_2x}{x_1+x_2}
-(x_1+x_2)\Big],\\
       T^{(a)}_3(0)
        &=& \frac{a_1M^2(\phi){b^\prime}^2}{3}
\int_0^\infty dx dx_1 dx_2 \frac{x x_2}{(x+a_2+a_1b^\prime)(x+1)}
\frac{1}{(x_1+x_2)^2}\exp\Big[-\frac{x_1x_2x}{x_1+x_2}
-(x_1+x_2)\Big].
\end{eqnarray}
Next, consider the integrals $T^{(a)}_4$ and $T^{(a)}_5$,
the evaluation of them
requires the results of the integrals $L_1$ and $L_2$.
$L_1$ is ultraviolet convergent and it reads for $k=0$ as
\begin{equation}
        L_1(p,0) = \int\frac{d^4k^\prime}{(2\pi)^4}
        {\cal{D}}^2(k^\prime){\cal{D}}(p+k^\prime).
\end{equation}
We go to the Euclidean space and after parameterizing the denominators
using Schwinger's method we integrate over $k^\prime$. Then we obtain
\begin{equation}
L_1(p_E,0) =\frac{1}{16\pi^2}\int_0^\infty d\alpha_1 d\alpha_2
\frac{\alpha_1}{(\alpha_1+\alpha_2)^2}
\exp\Big[-\frac{\alpha_1\alpha_2}{\alpha_1+\alpha_2}p_E^2
        -(\alpha_1+\alpha_2)M^2(\phi)\Big].
\end{equation}
Proceeding with the similar method $L_2$ has been obtained for $k=0$ as
\begin{equation}
L_2(p_E,0)=-\frac{i}{16\pi^2}\int_0^\infty d\alpha_1 d\alpha_2
\frac{\alpha_1\alpha_2}{(\alpha_1+\alpha_2)^2}
        \exp\left[-\frac{\alpha_1\alpha_2}{\alpha_1+\alpha_2}p_E^2
        -(\alpha_1+\alpha_2)M^2(\phi)\right].
\end{equation}
Upon substitution of $L_1$ and $L_2$ into $T^{(a)}_4$ and
$T^{(a)}_5$ respectively,
we perform integrations over $q_E$ and $r_E$ and obtain them for $k=0$ as
\begin{eqnarray}
	T^{(a)}_4(0) &=& \frac{2M^2(\phi){b^\prime}^2}{3}
\int_0^\infty dx dx_1\cdots dx_7
\frac{x x_1 (x+a_2)^2}{(x+a_2+a_1 b^\prime)^2}
        \frac{1}{(1+x)\eta_3^2}\exp[-\eta_4x-(x_1+\cdots+x_7)],\\
	T^{(a)}_5(0) &=& \frac{M^2(\phi){b^\prime}^2}{6}
\int_0^\infty dx dx_1\cdots dx_7
\frac{x x_1 x_2 (x+b)^2}{(x+b+a b^\prime)^2\eta_3^2}
\exp[-\eta_4x-(x_1+\cdots+x_7)],
\end{eqnarray}
where
\begin{eqnarray}
\eta_3 &=& (x_1+x_2)\{(x_6+x_7)(x_3+x_4+x_5)+x_3(x_4+x_5)\},\\
\eta_4 &=& \frac{x_1x_2}{x_1+x_2}
+\frac{x_3x_5x_6+x_4x_5x_6+x_4x_5x_7+x_3x_4x_5+x_6x_7(x_3+x_4+x_5)
        +x_3x_4x_7}
{(x_6+x_7)(x_3+x_4+x_5)+x_3(x_4+x_5)}.
\end{eqnarray}
Our numerical evaluation gives
\begin{equation}
        \frac{\partial V_{2}^{(a)}}{\partial G(k)}
\Big{|}_{G={\cal{D}}}
        = - \frac{M^2(\phi){b^\prime}^2}{6}
        \exp[b_1{b^\prime}^2-b_2{b^\prime}+b_3]
        +0(\frac{k^2\phi^2}{M^4(\phi)}),
\end{equation}
where $b_1=0.04482\pm 0.00286$,
$b_2=0.39237\pm 0.01053$ and
$b_3=2.91975\pm 0.00716$. It is to be mentioned here that
the leading term is of the order of $\phi^2/M^2(\phi)$ and it is
suppressed in the non-leading order by a factor of
$k^2/M^2(\phi)$.

\section{Evaluation of
$\frac{\partial V_{2}^{(b)}}{\partial G(k)}
\Big{|}_{G={\cal{D}}}$
and $\frac{\partial V_{2}^{(c)}}{\partial G(k)}
\Big{|}_{G={\cal{D}}}$}
\label{dGamma2bc}
\begin{equation}
\frac{\partial V_{2}^{(b)}}{\partial G(k)}
\Big{|}_{G={\cal{D}}}
        = - \sum_{i=1}^3 T^{(b)}_i(k),
\label{Gamma2bbyGk1}
\end{equation}
where,
\begin{eqnarray}
        T^{(b)}_1(k) &=& \frac{i\lambda^2}{12}\int\frac{d^4p}{(2\pi)^4}
        \frac{(I(p)-\frac{\lambda^2\phi^2}{2}J(p)){\cal{D}}(p-k)}
{1-\frac{i\lambda^3\phi^2}{4}J(p)+\frac{i\lambda}{2}I(p)},\\
        T^{(b)}_2(k)&=& \frac{i\lambda^2}{12}\int\frac{d^4p}{(2\pi)^4}
        \frac{I(p){\cal{D}}(p-k)}{(1-\frac{i\lambda^3\phi^2}{4}J(p)
+\frac{i\lambda}{2}I(p))^2},\\
        T^{(b)}_3(k) &=&-\frac{i\lambda^4\phi^2}{48}\int\frac{d^4p}{(2\pi)^4}
\frac{I(p)}{(1-\frac{i\lambda^3\phi^2}{4}J(p)
+\frac{i\lambda}{2}I(p))^2}
        (4{\cal{D}}(p-k)L_{1}(p,k)+L_{2}(p,k)).
\end{eqnarray}
Consider the integral $T^{(b)}_1$ which is written as
\begin{equation}
        T^{(b)}_1(k) = \frac{\lambda}{6}\int\frac{d^4p}{(2\pi)^4}
{\cal{D}}(p-k)
        \Big\{1
-\frac{2}{\lambda(iI(p)-\frac{\lambda^2\phi^2}{2}iJ(p)+\frac{2}{\lambda})}
        \Big\}
\end{equation}
Since,
\begin{eqnarray}
        I(p) &=& -\frac{i}{16\pi^2\epsilon}+0(1),\\
        J(p) &=& -\frac{ia_1}{256\pi^4}\frac{1}{p^2-a_2M^2(\phi)} 
        + 0(\epsilon)
\end{eqnarray}
we obtain
\begin{equation}
        \frac{1}{iI(p)-\frac{\lambda^2\phi^2}{2}iJ(p)+\frac{2}{\lambda})}
= 16\pi^2\epsilon + 0(\epsilon^2).
\end{equation}
Therefore,
\begin{equation}
        T^{(b)}_1(k)=\frac{\lambda}{6}
        \left(1-\frac{32\pi^2}{\lambda}\epsilon + 0(\epsilon^2)\right)
        \int\frac{d^4p}{(2\pi)^4}G(p-k).
\end{equation}
Similarly, considering the dependence of $I$ and $J$ on $\epsilon$ we obtain
$\frac{i\lambda}{2}I(p)(1-\frac{i\lambda^3\phi^2}{4}J(p)
+\frac{i\lambda}{2}I(p))^{-2}
=\frac{32\pi^2}{\lambda}\epsilon + 0(\epsilon^2)$. Then we obtain
$T^{(b)}_2$ and $T^{(b)}_3$ as
\begin{eqnarray}
        T^{(b)}_2(k) &=& \frac{16\pi^2}{3}\epsilon
        \int\frac{d^4p}{(2\pi)^4}{\cal{D}}(p-k),\\
        T^{(b)}_3(k) &=& -\frac{4\pi^2\lambda^2\phi^2}{3}\epsilon
\int\frac{d^4p}{(2\pi)^4}
        (4{\cal{D}}(p-k)L_{1}(p,k)+L_{2}(p,k)).
\end{eqnarray}
Consider the first integral in $T^{(b)}_3$:
\begin{equation}
        \int\frac{d^4p}{(2\pi)^4}{\cal{D}}(p-k)L_{1}(p,k)
        = \int\frac{d^4p}{(2\pi)^4}\frac{d^4k^\prime}{(2\pi)^4}
        {\cal{D}}(p-k){\cal{D}}(k^\prime){\cal{D}}(k+k^\prime)
        {\cal{D}}(p-k-k^\prime).
\end{equation}
We replace $p$ by $p+k$ on the right hand side and
obtain
\begin{equation}
\int\frac{d^4p}{(2\pi)^4}G(p-k)L_{1}(p,k)
= \int\frac{d^4k^\prime}{(2\pi)^4}
        {\cal{D}}(k^\prime){\cal{D}}(k+k^\prime)I(k^\prime).
\end{equation}
Next, consider the second integral in $T^{(b)}_3$:
\begin{equation}
        \int\frac{d^4p}{(2\pi)^4}L_{2}(p,k)
        =\int\frac{d^4p}{(2\pi)^4}\frac{d^4k^\prime}{(2\pi)^4}
        {\cal{D}}(k^\prime){\cal{D}}(p+k^\prime)
        {\cal{D}}(k+k^\prime){\cal{D}}(p+k+k^\prime).
\end{equation}
Now, we replace $p$ by $p-k^\prime$ on the right hand side and obtain
\begin{equation}
\int\frac{d^4p}{(2\pi)^4}L_{2}(p,k) = I^2(k).
\end{equation}
Then upon substitutions of $T^{(b)}_i$s into Eq.(\ref{Gamma2bbyGk1})
we obtain
\begin{equation}
\frac{\partial V_{2}^{(b)}}{\partial G(k)}
\Big{|}_{G={\cal{D}}}
        = -\frac{\lambda}{6} \int\frac{d^4p}{(2\pi)^4}{\cal{D}}(p)
+ \frac{4\pi^2\lambda^2\phi^2}{3}\epsilon
\Big\{I^2(k)
        +4\int\frac{d^4k^\prime}{(2\pi)^4}
        {\cal{D}}(k^\prime){\cal{D}}(k+k^\prime)I(k^\prime)\Big\}.
\label{Gamma2bbyGk2}
\end{equation}
The first integral evaluated in the dimension
$d=4-2\epsilon$ ($\epsilon>0$) gives
\begin{equation}
        \int\frac{d^4p}{(2\pi)^4}{\cal{D}}(p) =
        \frac{i}{16\pi^4}(4\pi^2\mu^2)^\epsilon
        \int\frac{d^dp}{p^2-M^2(\phi)}
        =-\frac{M^2(\phi)}{16\pi^2}\left\{\frac{1}{\epsilon}+1-\gamma_E
        -\ln\left(\frac{M^2(\phi)}{4\pi\mu^2}\right)\right\}.
\label{Gamma2bint1i}
\end{equation}
Consider the second integral
\begin{eqnarray}
& &\int\frac{d^4k^\prime}{(2\pi)^4}
        {\cal{D}}(k^\prime){\cal{D}}(k+k^\prime)I(k^\prime)\nonumber\\
	&=& \frac{i\Gamma(1+\epsilon)}{16\pi^2\epsilon}
(4\pi\mu^2)^\epsilon(-1)^\epsilon
\int_0^1 dx (x(1-x))^{-\epsilon}
        \int\frac{d^4k^\prime}{(2\pi)^4}
        \frac{({k^\prime}^2-\frac{M^2(\phi)}{x(1-x)})^{-\epsilon}}
        {({k^\prime}^2-M^2(\phi))
        ((k+k^\prime)^2-M^2(\phi))}\nonumber\\
        &=& \frac{i\Gamma(2+\epsilon)}{256\pi^6}(4\pi\mu^2)^{2\epsilon}
\int_0^1 dx\, dy\, dz\, (x(1-x))^{-\epsilon}\, y^{-1+\epsilon}
        \int
d^dk^\prime[-{k^\prime}^2-2zk.k^\prime
        +M^2(\phi)(1-y+\frac{y}{x(1-x)})
        -k^2z]^{-2-\epsilon},
\end{eqnarray}
where the Feynman's parametrization and the dimensional 
regularization have been used.
Performing the integration over $k^\prime$ we obtain
\begin{eqnarray}
& &\int\frac{d^4k^\prime}{(2\pi)^4}
        {\cal{D}}(k^\prime){\cal{D}}(k+k^\prime)I(k^\prime)\nonumber\\
&=& -\frac{\Gamma(2\epsilon)}{256\pi^4}(4\pi\mu^2)^{2\epsilon}
\int_0^1 dx (x(1-x))^{-\epsilon}\int_0^1 dz\int_0^1 dy
y^{-1+\epsilon}
        \left[M^2(\phi)(1-y+\frac{y}{x(1-x)})-k^2z(1-z)\right]^{-2\epsilon}.
\end{eqnarray}
Now perform the integration by parts with respect to $y$ taking
$y^{-1+\epsilon}$
as the second function.
\begin{eqnarray}
        & &\epsilon\int\frac{d^4k^\prime}{(2\pi)^4}
        {\cal{D}}(k^\prime){\cal{D}}(k+k^\prime)I(k^\prime)
= -\frac{\Gamma(2\epsilon)}{256\pi^4}
\int_0^1 dx\, (x(1-x))^{\epsilon}\int_0^1 dz
        \left[\frac{M^2(\phi)
        -k^2z(1-z)x(1-x)}{4\pi\mu^2}\right]^{-2\epsilon}\nonumber\\
        & & -\frac{\Gamma(1+2\epsilon)M^2(\phi)}{256\pi^4}
\int_0^1 dx\, (x(1-x))^{\epsilon}\, (1-x(1-x))\int_0^1 dz\nonumber\\
        & &\times\int_0^1 dy
        \frac{y^\epsilon}{\left[M^2(\phi)(y+x(1-x)(1-y))-k^2z(1-z)x(1-x)
\right]^{1+2\epsilon}}.
\end{eqnarray}
We expand the right hand side about $\epsilon =0$ to order $1$ and obtain
\begin{eqnarray}
& &\epsilon\int\frac{d^4k^\prime}{(2\pi)^4}
        {\cal{D}}(k^\prime){\cal{D}}(k+k^\prime)I(k^\prime)
= -\frac{1}{512\pi^4}
\Big[\frac{1}{\epsilon} - 2\gamma_E - 2
        -2\int_0^1 dx dz\ln\Big(\frac{M^2(\phi)-k^2z(1-z)x(1-x)}{4\pi\mu^2}
\Big)\Big]\nonumber\\
        &-&\frac{M^2(\phi)}{256\pi^4}\int_0^1 dx\, dy\, dz
\frac{1-x(1-x)}
        {M^2(\phi)(y+x(1-x)(1-y))-k^2z(1-z)x(1-x)} 
        + 0(\epsilon).
\label{Gamma2bint2}
\end{eqnarray}
Next, consider the term
\begin{eqnarray}
        \epsilon I^2(k)
        &=& - \frac{\Gamma^2(1+\epsilon)}{256\pi^4\epsilon}
        \Big(\int_0^1 dx \left(\frac{M^2(\phi)-k^2x(1-x)}
{4\pi\mu^2}\right)^{-2\epsilon}\Big)^2\nonumber\\
        &=&-\frac{1}{256\pi^4}\Big[\frac{1}{\epsilon}-2\gamma_E
        -2\int_0^1 dx \ln\Big(\frac{M^2(\phi)-k^2x(1-x)}{4\pi\mu^2}\Big)
+ 0(\epsilon)\Big].
\label{Gamma2bterm}
\end{eqnarray}
 Upon substitutions of the results of Eq.(\ref{Gamma2bint1i}),
 (\ref{Gamma2bint2}) and (\ref{Gamma2bterm})
 into (\ref{Gamma2bbyGk2}) we obtain
\begin{eqnarray}
        & &\frac{\partial V_{2}^{(b)}}{\partial G(k)}
        \Big{|}_{G={\cal{D}}}
        = + \frac{\lambda}{96\pi^2\epsilon}\Big(M^2(\phi)
        -\frac{3}{2}\lambda\phi^2\Big)
        +\frac{\lambda M^2(\phi)}{96\pi^2}\Big\{1-\gamma_E
        -\ln\Big(\frac{M^2(\phi)}{4\pi\mu^2}\Big)\Big\}\nonumber\\
& &+\frac{\lambda^2\phi^2}{96\pi^2}\Big\{+ 2 + 3\gamma_E
        + \int_0^1 dx  \ln\Big(\frac{M^2(\phi)-k^2x(1-x)}{4\pi\mu^2}\Big)
        + 2\int_0^1dx\, dz 
	\ln\Big(\frac{M^2(\phi)-k^2z(1-z)x(1-x)}{4\pi\mu^2}
\Big)\nonumber\\
        & &- 2 M^2(\phi)\int_0^1 dx\, dy\, dz
\frac{1-x(1-x)}
        {M^2(\phi)(y+x(1-x)(1-y))-k^2z(1-z)x(1-x)}\Big\} + 0(\epsilon).
\end{eqnarray}
We expand the right hand side to order $k^2/M^2(\phi)$ and perform the
integrations over the parameters. The result is
\begin{eqnarray}
\frac{\partial V_{2}^{(b)}}{\partial G(k)}
\Big{|}_{G={\cal{D}}}
	& =& \frac{\lambda}{96\pi^2\epsilon}\Big(M^2(\phi)
        -\frac{3}{2}\lambda\phi^2\Big)
- \frac{\lambda^2\phi^2}{96\pi^2}\Big\{2-3\gamma_E
        -3\ln\Big(\frac{M^2(\phi)}{4\pi\mu^2}\Big)
        +\frac{1}{2}\frac{k^2}{M^2(\phi)}\Big\}\nonumber\\ 
        & &+\frac{\lambda M^2(\phi)}{96\pi^2}\Big\{1-\gamma_E
        -\ln\Big(\frac{M^2(\phi)}{4\pi\mu^2}\Big)\Big\} 
        + 0(\frac{k^4}{M^4(\phi)}).
\end{eqnarray}

We use Eq.(\ref{Gamma2bint1i}) to obtain
\begin{equation}
\frac{\partial V_{2}^{(c)}}{\partial G(k)}
\Big{|}_{G={\cal{D}}}
        = \frac{\lambda}{4}\int\frac{d^4p}{(2\pi)^4} {\cal{D}}(p)
        = - \frac{\lambda M^2(\phi)}{64\pi^2}\left\{\frac{1}{\epsilon}
        +1-\gamma_E -\ln\left(\frac{M^2(\phi)}{4\pi\mu^2}\right)\right\}.
\end{equation}

\section{Evaluation of integrals in $V_{eff}$}
\label{veffintegrals}

Consider the integral
\begin{equation}
        I_1 =-\frac{i}{2}\int\frac{d^4p}{(2\pi)^4}\ln G^{-1}(p)
        = -\frac{M_1^4(\phi)}{64\pi^2Z^2}\Big\{\frac{1}{\epsilon}
+\frac{3}{2}-\gamma_E
        -\ln\Big(\frac{M_1^2(\phi)}{4\pi\mu^2Z}\Big) + 0(\epsilon)\Big\}
\label{I1}
\end{equation}
where the result is arbitrary up to an additive constant independent of
$M_1(\phi)$.

Next, consider the integral
\begin{equation}
I_2=-\frac{i}{2}\int\frac{d^4p}{(2\pi)^4}G^{-1}_{ct}G(p).
\end{equation}
Since, $iG^{-1}_{ct}$ is independent of $p$  we regularize the integral
in $d$ dimension and after carrying out integration over $p$ we obtain
\begin{equation}
	I_2 = \frac{M^2_1(\phi)(-iG^{-1}_{ct})}{32\pi^2Z^2}\Gamma(-1+\epsilon)
        \Big(\frac{M^2_1(\phi)}{4\pi\mu^2Z}\Big)^{-\epsilon}.
\end{equation}
Now the use of Eq.(\ref{Ginvct}) with
expansion of the result to order $\epsilon^2$, we obtain
\begin{eqnarray}
            I_2 &=& -\frac{M_1^2(\phi)}{32\pi^2Z^2}
\Big[\frac{\lambda}{96\pi^2\epsilon^2}
\Big(m^2+\frac{7}{2}\lambda\phi^2\Big)
+\frac{\lambda}{96\pi^2\epsilon}\Big(m^2+\frac{7}{2}\lambda\phi^2\Big)
\Big(1-\gamma_E-\ln \Big(\frac{M_1^2(\phi)}{4\pi\mu^2 Z}\Big)\Big)\nonumber\\
	& &+\frac{\lambda m^2}{96\pi^2\epsilon}\Big\{1-\gamma_E
	-\ln\Big(\frac{m^2}{4\pi\mu^2}\Big)\Big\}
        +\frac{\lambda}{192\pi^2}\Big(m^2+\frac{7}{2}\lambda\phi^2\Big)
\Big\{1+\frac{\pi^2}{6}+\Big(1-\gamma_E
-\ln\Big(\frac{M_1^2(\phi)}{4\pi\mu^2Z}\Big)\Big)^2\Big\}\nonumber\\
	 & &+ \frac{\lambda m^2}{96\pi^2}\Big(1-\gamma_E
-\ln \Big(\frac{m^2}{4\pi\mu^2}\Big)\Big)
\Big(1-\gamma_E-\ln \Big(\frac{M_1^2(\phi)}{4\pi\mu^2 Z}\Big)\Big)\Big],
\label{I2}
\end{eqnarray}

The integral
\begin{equation}
I_3 = -\frac{i}{2}\int\frac{d^4p}{(2\pi)^4}
{\cal{D}}^{-1}(\phi, p)G(p),
\end{equation}
is ultraviolet divergent. We carry out the $p$ integration
in $d$ dimension and obtain the result to order $1$ as
\begin{equation}
        I_3 = \frac{M_1^2(M_1^2-Z M^2(\phi))}{32\pi^2 Z^3}
\Big[\frac{1}{\epsilon}+1-\gamma_E
-\ln\Big(\frac{M^2_1}{4\pi\mu^2Z}\Big)
+ 0(\epsilon)\Big].
\label{I3}
\end{equation}

The integral
\begin{equation}
I_4 = i \int\frac{d^4p}{(2\pi)^4}\ln S^{-1}(p)
	= \frac{M_f^4(\phi)}{32\pi^2}
\Big[\frac{1}{\epsilon} -\frac{1}{2}-\gamma_E
	-\ln\Big(\frac{M^2_f(\phi)}{4\pi\mu^2}\Big)\Big].
\label{I4}
\end{equation}

Consider the integral in Eq.(\ref{Gamma2a}) which is rewritten as
\begin{equation}
V^{(a)}_2(\phi, G)=
\frac{\lambda^3\phi^2}{24}
\int \frac{d^4p}{(2\pi)^4}
\Big[J(p) - \frac{J(p)}{1-\frac{i\lambda^3\phi^2}{4}J(p)}\Big].
\label{Gamma2a1}
\end{equation}
The Eq.(\ref{Jx}) gives
\begin{equation}
        J(p) = -\frac{i}{256\pi^4Z^5}
\Big[\frac{a_1l_\epsilon}{p^2-\bar{M}_1^2a_2}
        -\frac{a_3\epsilon}{p^2-\bar{M}_1^2a_4}\Big],
        \label{Jp1}
\end{equation}
where $l_\epsilon = 1-2\epsilon\ln(\bar{M}_1^2/4\pi\mu^2)$
and $\bar{M}_1^2= M_1^2/Z$.
We use this form of $J(p)$ to obtain
\begin{eqnarray}
\frac{J(p)}{1-\frac{i\lambda^3\phi^2}{4}J(p)}
        &=& -\frac{i}{256\pi^4(m_+^2-m_-^2)Z^5}
\Big[
        \frac{m_+^2(l_\epsilon a_1-\epsilon a_3)
        -\bar{M}_1^2(l_\epsilon a_1a_4-\epsilon a_2a_3)}
{p^2-m_+^2}\nonumber\\
        & & -\frac{m_-^2(l_\epsilon a_1-\epsilon a_3)
        -\bar{M}_1^2(l_\epsilon a_1a_4-\epsilon a_2a_3)}
{p^2-m_-^2}\Big],
\label{Jp2}
\end{eqnarray}
\begin{eqnarray}
        m_\pm^2 &=& \frac{1}{2}((a_4+a_2)\bar{M}_1^2
+b^{\prime\prime}(l_\epsilon a_1-\epsilon a_3))
        \pm\frac{1}{2}\sqrt{((a_4-a_2)\bar{M}_1^2
-b^{\prime\prime}(l_\epsilon a_1-\epsilon a_3))^2
        -4\epsilon a_3(a_4-a_2)b^{\prime\prime}\bar{M}_1^2},\\
b^{\prime\prime} &=& \frac{\lambda^3\phi^2}{1024\pi^4Z^5}.
\end{eqnarray}
We substitute Eq.(\ref{Jp1}) and (\ref{Jp2}) in Eq.(\ref{Gamma2a1})
and regularize the integral in $d$ dimension. Then, we perform the
integration over $p$ and obtain the result to order $1$ as
\begin{eqnarray}
& &V^{(a)}_2(\phi, G) =
        - \frac{a_1^2{b^{\prime\prime}}^2}{96\pi^2\epsilon}
- \frac{a_1b^{\prime\prime}}{96\pi^2 Z}
        \Big[M_1^2(\phi)a_2\ln{a_2}+a_1(1-\gamma_E)Zb^{\prime\prime}
        -\Big(a_2M_1^2(\phi)+a_1Zb^{\prime\prime}\Big)
        \ln\Big(\frac{a_2M_1^2(\phi)+a_1Zb^{\prime\prime}}{4\pi\mu^2Z}
	\Big)\nonumber\\
        & &-\frac{2a_3Zb^{\prime\prime}}
	{((a_4-a_2)M_1^2(\phi)-a_1Zb^{\prime\prime})^2}
        \Big\{(a_4-a_2)^2M_1^4(\phi)
	-a_1(2a_4-3a_2)M_1^2(\phi)Zb^{\prime\prime}
+2a_1^2Z^2{b^{\prime\prime}}^2\Big\}\nonumber\\
        & &+\frac{1}{(a_4-a_2)M_1^2(\phi)-a_1Zb^{\prime\prime}}
        \Big\{a_2(a_4-a_2)M_1^4(\phi)
        -a_1(4a_4-3a_2)M_1^2(\phi)Zb^{\prime\prime}
	+ 4a_1^2Z^2{b^{\prime\prime}}^2\Big\} 
        \ln\Big(\frac{M_1^2(\phi)}{4\pi\mu^2Z}\Big)\Big].
\label{Gamma2a2}
\end{eqnarray}

Consider the integral in Eq.(\ref{Gamma2b}) which is rewritten as
\begin{equation}
 V^{(b)}_2(\phi, G) =
- \frac{\lambda}{12}\int \frac{d^4p}{(2\pi)^4}
\Big[I(p)+\frac{2i}{\lambda}
        +\frac{(iJ(p))(i\lambda\phi^2)}
{iI(p)-\frac{\lambda^2\phi^2}{2}iJ(p)+\frac{2}{\lambda}}
-\frac{4i/\lambda^2}{iI(p)-\frac{\lambda^2\phi^2}{2}iJ(p)+\frac{2}{\lambda}}
\Big].
\end{equation}
The second term after $p$ integration will give zero. Now consider
the term
\begin{equation}
\Big[iI(p)-\frac{\lambda^2\phi^2}{2}iJ(p)+\frac{2}{\lambda}\Big]^{-1}
        =\Big[\frac{\Gamma(1+\epsilon)}{16\pi^2Z^2\epsilon}
        \Big(\frac{\bar{M}_1^2}{4\pi\mu^2}\Big)^{-\epsilon} \hat{I}(p)
        -\frac{a_1\lambda^2\phi^2}{512\pi^4Z^5}\frac{1}{p^2-\bar{M}_1^2a_2}
        +\frac{2}{\lambda}
+0(\epsilon)\Big]^{-1}
        = 16\pi^2Z^2\epsilon + 0(\epsilon^2),
\end{equation}
where we have used the fact that $\hat{I}(p)=1+0(\epsilon)$.
Similarly, we obtain
\begin{equation}
\frac{iJ(p)}{iI(p)-\frac{\lambda^2\phi^2}{2}iJ(p)+\frac{2}{\lambda}}
        =\frac{a_1\epsilon}{16\pi^2 Z^3}\frac{1}{p^2-\bar{M}_1^2a_2}
        +0(\epsilon^2).
\end{equation}
The results give
\begin{equation}
        V^{(b)}_2(\phi, G) =
- \frac{\lambda}{12}\int \frac{d^4p}{(2\pi)^4}
        \Big[I(p) + \frac{ia_1\lambda\phi^2\epsilon}{16\pi^2Z^3}
        \frac{1}{p^2-\bar{M}_1^2a_2}
-\frac{64\pi^2i\epsilon}{\lambda^2}
        + 0(\epsilon^2)\Big]
\label{Gamma2b1}
\end{equation}
The third term is $p$ independent and it gives zero after performing
$p$ integration. The first and the second terms are ultraviolet
divergent and those are to be evaluated after regularizing them
in $d$ dimensions. Consider the first integral which takes the
following form after the use of Eq.(\ref{Ip1}):
\begin{eqnarray}
	 \frac{\lambda}{12}\int \frac{d^4p}{(2\pi)^4} I(p)
	&=&-\frac{i\lambda\Gamma(1+\epsilon)}{192\pi^2Z^2\epsilon}
        \frac{(4\pi\mu^2)^{2\epsilon}\pi^\epsilon}{16\pi^4}
\int_0^1 dx (x(1-x))^{-\epsilon}
        \int\frac{d^dp}{\Big(\frac{\bar{M}_1^2}{x(1-x)}
        -p^2\Big)^\epsilon}\nonumber\\
	&=& \frac{\lambda \bar{M}_1^4\Gamma(-2+2\epsilon)}{3072\pi^4 Z^2}
        \Big(\frac{\bar{M}_1^2}{4\pi\mu^2}\Big)^{-2\epsilon}
        \int_0^1 dx (x(1-x))^{-2+\epsilon}
        = \frac{\lambda \bar{M}_1^4}{3072\pi^4 Z^2}
        \Big(\frac{\bar{M}_1^2}{4\pi\mu^2}\Big)^{-2\epsilon}
(\Gamma(-1+\epsilon))^2\nonumber\\
        &=& \frac{\lambda \bar{M}_1^4}{3072\pi^4Z^2\epsilon^2}\Big[1
        +2\epsilon\Big\{1-\gamma_E-\ln\Big(\frac{\bar{M}_1^2}
        {4\pi\mu^2}\Big)\Big\}
+\epsilon^2\Big\{3-4\gamma_E+2\gamma_E^2+\frac{\pi^2}{6}
        -4(1-\gamma_E)\ln\Big(\frac{\bar{M}_1^2}{4\pi\mu^2}\Big)\nonumber\\
	&+&2\ln^2\Big(\frac{\bar{M}_1^2}{4\pi\mu^2}\Big)\Big\} 
        + 0(\epsilon^3)\Big]
\end{eqnarray}
The second integral gives
\begin{equation}
        \frac{ia_1\lambda^2\phi^2\epsilon}{192\pi^2Z^3}
        \int\frac{d^4p}{(2\pi)^4}\frac{1}{p^2-\bar{M}_1^2a_2}
        = \frac{ia_1\lambda^2\phi^2\epsilon}{192\pi^2Z^3}
\frac{(4\pi^2\mu^2)^\epsilon}{16\pi^4}
        \int\frac{d^dp}{p^2-\bar{M}_1^2a_2}
        = -\frac{a_1a_2\lambda^2\phi^2 \bar{M}_1^2}{3072\pi^4Z^3} 
        + 0(\epsilon).
\end{equation}
Upon substitution of these results into Eq.(\ref{Gamma2b1}), we
obtain
\begin{eqnarray}
	V^{(b)}_2(\phi, G) &=&
        -\frac{\lambda M_1^4(\phi)}{3072\pi^4Z^4\epsilon^2}
        - \frac{\lambda M_1^4(\phi)}{1536\pi^4Z^4\epsilon}
        \Big\{1-\gamma_E-\ln\Big(\frac{M_1^2(\phi)}{4\pi\mu^2Z}\Big)\Big\}
        - \frac{\lambda M_1^4(\phi)}{3072\pi^4Z^4}
        \Big\{2\Big(1-\gamma_E
	-\ln\Big(\frac{M_1^2(\phi)}{4\pi\mu^2Z}\Big)\Big)^2
        \nonumber\\
        & &+ \frac{\pi^2}{6}+1\Big\}
	+ \frac{a_1a_2\lambda^2\phi^2 M_1^2(\phi)}{3072\pi^4Z^4}
+ 0(\epsilon).
        \label{Gamma2b2}
\end{eqnarray}

\begin{eqnarray}
	V^{(c)}_2(\phi, G) &=& \frac{\lambda}{8}
        \left(\int\frac{d^4p}{(2\pi)^4} G(p)\right)^2
        = \frac{\lambda M_1^4(\phi)}{2048\pi^4 Z^4}
        \left(\frac{M_1^2(\phi)}{4\pi\mu^2 Z}\right)^{-2\epsilon}
\frac{\Gamma^2(1+\epsilon)}{\epsilon^2(1-\epsilon)^2}\nonumber\\
        &=& \frac{\lambda M_1^4(\phi)}{2048\pi^4 Z^4}
\Big[\frac{1}{\epsilon^2} + \frac{2}{\epsilon}
        \Big(1-\gamma_E-\ln\Big(\frac{M_1^2(\phi)}{4\pi\mu^2Z}\Big)\Big)
	+1+\frac{\pi^2}{6} 
        + 2\Big(1-\gamma_E-\ln\Big(\frac{M_1^2(\phi)}
	{4\pi\mu^2Z}\Big)\Big)^2\Big].
\label{Gamma2c2}
\end{eqnarray}


\begin{thebibliography}{00}

\bibitem{Jaffe1965}A. M. Jaffe, Divergence of perturbation theory for
		bosons, Comm. Math. Phys. 1 (1965) 127.

\bibitem{Dashen1974}R. F. Dashen, B. Hasslacher and A. Neveu, Nonperturbative 
	methods and extended-hadron models in field theory. II. Two-dimensional         models and extended hadrons, Phys. Rev. D 10 (1974) 4114.

\bibitem{Lipatov1977}L. N. Lipatov, Divergence of the perturbation-theory
	series and the quasi-classical theory, Sov. Phys. JETP 45 (1977) 216.

\bibitem{Coleman1973}S. Coleman and E. Weinberg, Radiative Corrections as 
	the Origin of Spontaneous Symmetry Breaking,
	Phys. Rev. D 7 (1973) 1888.	

\bibitem{Brezin1977}E. Brezin, J. C. Le Guillou and J. Zinn-Justin,
	Perturbation theory at large order. I. The $\phi^{2N}$ interaction,
	Phys. Rev. D 15 (1977) 1544.

\bibitem{Cornwall1974}J. M. Cornwall and R. Jackiw and E. Tomboulis,
	Effective action for composite operators,
	Phys. Rev. D 10 (1974) 2428.

\bibitem{Camelia1993}G. Amelino-Camelia and S-Young Pi, Self-consistent 
	improvement of the finite-temperature effective potential,
	Phys. Rev. D 47 (1993) 2356.

\bibitem{Camelia1996}G. Amelino-Camelia, On the CJT formalism in multi-field 
	theories, Nucl. Phys. B 476 (1996) 255.

\bibitem{Alford2008}M. G. Alford, M. Braby, and A. Schmitt, Critical  
	temperature for kaon condensation in color-flavor locked quark matter,
	J. Phys. G 35 (2008) 025002.

\bibitem{Andersen2009}J. O. Andersen and L. E. Leganger, Kaon condensation 
	in the color-flavor-locked phase of quark matter, the Goldstone theorem,        and the 2PI Hartree approximation,
	Nucl. Phys. A 828 (2009) 360.

\bibitem{Tran2008}H. P. Tran, V. Nguyen, T. A. Nguyen and V. H. Le, Kaon 
	condensation in the linear sigma model at finite density and 
	temperature, Phys. Rev. D 78 (2008) 105016.

\bibitem{Agodi1998}A.Agodi, G.Andronico, P.Cea, M.Consoli and L.Cosmai,
        The $\phi^4$ theory on the lattice: effective potential and triviality,
	Nuclear Physics B (Proc. Suppl.) 63A-C (1998) 637-639.

\bibitem{Chung1997}J.M.Chung and B.K.Chung, Three-loop renormalization of the 
	effective potential, Phys. Rev. D 56 (1997) 6509.

\bibitem{Ananos2006}G. N. J. Añaños, Scalar field theory at finite temperature
	in $D=2+1$, J. Math. Phys. 47 (2006) 012301.

\bibitem{Peskin and Schroeder} M. E. Peskin and D. V. Schroeder,
An Introduction to Quantum Field Theory, Addison-Wesley Publishing 
		Company, USA (1995).

\bibitem{Collins}John C. Collins,
Renormalization, Cambridge Monographs on Mathematical Physics,
		Cambridge University Press (1984).
\bibitem{Jackiw1974} R. Jackiw, Functional evaluation of effective 
	potential, Phys. Rev. D 9 (1974) 1686.

\bibitem{Raymond}P. Raymond, Field Theory: A Modern Primer,
	Addison-Wesley Publishing Company, USA (1990).


\end{thebibliography}
\end{document}